\documentclass[seceqn,dvips]{arxstspdf}
\usepackage{graphics}
\usepackage{flushend,stfloats}

\volume{25}
\issue{1}
\pubyear{2010}
\firstpage{88}
\lastpage{106}
\doi{10.1214/10-STS325}

\makeatletter
\newproclaim{BF}{Bootstrap filter (BF)}
\newproclaim{APF}{Auxiliary particle filter (APF)}
\newproclaim{SF}{Storvik's filter}
\newproclaim{LWF}{Liu and West's filter}
\newproclaim{PL}{Particle learning}
\newproclaim{example}{Example}
\newproclaim{PS}{Particle smoothing}
\newproclaim{MM}{Model monitoring}
\newproclaim{alg}{Algorithm}
\makeatother

\begin{document}
\begin{frontmatter}

\title{Particle Learning and Smoothing}
\runtitle{Particle Learning}

\begin{aug}
\author[a]{\fnms{Carlos M.} \snm{Carvalho}\ead[label=e1]{carlos.carvalho@chicagobooth.edu}},
\author[b]{\fnms{Michael S.} \snm{Johannes}\ead[label=e2]{mj335@columbia.edu}},
\author[c]{\fnms{Hedibert F.} \snm{Lopes}\ead[label=e3]{hlopes@chicagobooth.edu}\corref{}}
\and
\author[d]{\fnms{Nicholas G.} \snm{Polson}\ead[label=e4]{ngp@chicagobooth.edu}}
\runauthor{Carvalho, Johannes, Lopes and Polson}

\affiliation{University of Chicago Booth School of Business, Columbia
University, University of Chicago Booth School of Business and
University of Chicago Booth School of Business}

\address[a]{Carlos M. Carvalho is Assistant Professor of Econometrics
and Statistics, University of Chicago Booth School of Business, 5807 South Woodlawn Avenue,
Chicago, Illinois 60637, USA \printead{e1}.}
\address[b]{Michael Johannes is Roger F. Murray Associate Professor of
Finance, Graduate School of Business,
Columbia University, 3022 Broadway, Uris Hall 424, New York, NY 10027, USA \printead{e2}.}
\address[c]{Hedibert F. Lopes is Associate Professor of Econometrics
and Statistics, University of Chicago Booth School of Business, 5807 South Woodlawn
Avenue Chicago, Illinois 60637, USA \printead{e3}.}
\address[d]{Nicholas G. Polson is Professor of Econometrics and Statistics,
University of Chicago Booth School of Business, 5807 South Woodlawn Avenue Chicago,
Illinois 60637, USA \printead{e4}.}

\end{aug}

%
\begin{abstract}
Particle learning (PL) provides state filtering, sequential parameter
learning and smoothing in a general class of state space models. Our
approach extends existing particle methods by incorporating the
estimation of static parameters via a fully-adapted filter that
utilizes conditional sufficient statistics for parameters and/or states
as particles. State smoothing in the presence of parameter uncertainty
is also solved as a by-product of PL. In a number of examples, we show
that PL outperforms existing particle filtering alternatives and proves
to be a competitor to MCMC.
\end{abstract}

%

\end{frontmatter}

\section{Introduction}

There are two statistical inference problems associated with state
space models. The first is sequential
state filtering and parameter learning, which is characterized by the
joint posterior distribution of parameters
and states at each point in time. The second is state smoothing, which
is characterized by the distribution of
the states, conditional on all available data, marginalizing out the
unknown parameters.

In linear Gaussian models, assuming knowledge about the system
parameters, the Kalman filter (Kal\-man, \citeyear{K1960}) provides the standard
analytical recursions for filtering and smoothing (West and Harrison,
\citeyear{WH1997}). For more general model specifications, conditional on
parameters, it is common to use sequential Monte Carlo methods known as
particle filters to approximate the sequence of filtering distributions
(see Doucet, de Freitas and Gordon, \citeyear{DFG2001} and Capp\'e,
Godsill and Moulines,
\citeyear{CGM2007}). As for\break smoothing, the posterior for states is typically
approximated via Markov chain Monte Carlo (MCMC) methods as developed
by Carlin, Polson and Stoffer (\citeyear{CPS1992}), Carter and Kohn
(\citeyear{CK1994}) and
Fr\"uhwirth-Schnatter (\citeyear{FS1994}).

In this paper we propose a new approach, called particle learning (PL),
for approximating the sequence of filtering and smoothing distributions
in light of parameter uncertainty for a wide class of state space
models. The central idea behind PL is the creation of a particle
algorithm that directly samples from the particle approximation to the
joint posterior distribution of states and conditional sufficient
statistics for fixed parameters in a fully-adapted resample--propagate
framework.

In terms of models, we consider Gaussian Dynamic Linear Models (DLMs)
and conditionally Gaussian (CDLMs). In these class of models, PL is
defined over both state and parameter sufficient statistics. This is a
generalization of the mixture Kalman filter (MKF) of Chen and Liu
(\citeyear{CL2000}) that allows for parameter learning. Additionally,
we show that
PL can handle nonlinearities in the state evolutions, dramatically
widening the class of models that MKF particle methods apply to.
Finally, we extend the smoothing results of Godsill, Doucet and West
(\citeyear{GDW2004}) to sequential parameter learning and to all the models
considered.

In a series of simulation studies, we provide significant empirical
evidence that PL dominates the standard particle filtering alternatives
in terms of estimation accuracy and that it can be seen as a true
competitor to MCMC strategies.

The paper starts in Section \ref{sec2}, with a brief review of the most popular
particle filters that represent the building blocks for the development
of PL in Section \ref{sec:general}. Section \ref{sec:CDLM} in entirely
dedicated to the application
of PL to CDLMs followed by possible extensions to nonlinear
alternatives in Section \ref{sec:nonlin}. Section \ref{sec6} presents a
series of experiments
benchmarking the performance of PL and highlighting its advantages over
currently used alternatives.

\section{Particle Filtering in State~Space~Models}\label{sec2}
Consider a general state space model defined by the observation and
evolution equations:
\begin{eqnarray*}
y_{t+1} & \sim& p( y_{t+1}| x_{t+1},\theta), \\
x_{t+1} & \sim& p(x_{t+1} | x_{t}, \theta),
\end{eqnarray*}
with initial state distribution $p(x_0|\theta)$ and prior $p(\theta)$.
In the above notation, states at time $t$ are represented by $x_t$
while the static parameters are denoted by $\theta$.
The sequential state filtering and parameter learning problem is solved
by the sequence of joint posterior distributions, $p(x_t,\theta|y^t)$,
where $y^t=(y_1,\dots,y_t)$ is the set of observations up to time $t$.

Particle methods use a discrete representation of $p(x_t,\theta|y^t)$ via
\[
p^N( x_t , \theta| y^t ) = \frac{1}{N} \sum_{i=1}^N \delta_{ ( x_t ,
\theta)^{(i)} },
\]
where $(x_t,\theta)^{(i)}$ is the state and parameter particle vector
and $\delta_{(\cdot)}$ is the Dirac measure,
representing the distribution degenerate at the $N$ particles.
Given this approximation, the key problem is how to
sample from this joint distribution sequentially as new data arrives.
This step is complicated because the state's propagation depends on the
parameters, and vice versa.
To circumvent the codependence in a joint draw, it is common to use
proposal distributions in a sequence of importance sampling steps.
We now review the main approaches of this general sequential Monte
Carlo strategy first for pure filtering and then with parameter learning.

\subsection{Pure Filtering Review}
\label{sec:review}
We start by considering the pure filtering problem, where it is assumed
that the set of parameters $\theta$ is known. Although less relevant in
many areas of application, this is the traditional engineering
application where both the Kalman filter and original particle filters
were developed.

\subsubsection*{The bootstrap filter}
In what can be considered the seminal work in the particle filtering
literature, Gordon, Salmond and Smith (\citeyear{GSS1993}) developed a strategy based on a
sequence of importance sampling steps where the proposal is defined by
the prior for the states.
This algorithm uses the following representation of the filtering
density:
\[
p( x_{t+1}|y^{t+1}) \propto p(y_{t+1}|x_{t+1})
p( x_{t+1}|y^{t}),
\]
where the state predictive is
\[
p(x_{t+1}|y^{t}) =\int p( x_{t+1}|x_{t}) p(x_{t}|y^{t})\, dx_{t}.
\]
Starting with a particle approximation of $p(x_t|y^t),$ draws from
$p(x_{t+1}|y^t)$ are obtained by propagating the particles forward via
the evolution equation $p(x_{t+1}|x_t)$, leading to importance sampling
weights that are proportional to like likelihood $p(y_{t+1}|x_{t+1})$.
The bootstrap filter can be summarized by the following:
\begin{BF*}
\begin{longlist}
\item[\textit{Step 1} (Propagate).]  $\{x_{t}^{(i)}\}_{i=1}^N$ to $\{{\tilde
x}_{t+1}^{(i)}\}_{i=1}^N$ via\break $p(x_{t+1}|x_{t})$.
\item[\textit{Step 2} (Resample).] $\{x_{t+1}^{(i)}\}_{i=1}^N$ from $\{
{\tilde x}_{t+1}^{(i)}\}_{i=1}^N$\vspace*{1.5pt} with weights
$w_{t+1}^{(i)} \propto p(y_{t+1}|{\tilde x}_{t+1}^{(i)})$.
\end{longlist}
\end{BF*}

Resampling in the second stage is an optional step, as any quantity of
interest could be computed more accurately by the use of the particles
and its associated weights. Resampling has been used as a way to avoid
the decay in the particle approximation and we refer the reader to Liu
and Chen (\citeyear{LC1998}) for a careful discussion of its merits. Throughout our
work we describe all filters with a resampling step, as this is the
central idea to our particle learning strategy introduced below.
Notice, therefore, that we call BF a \textit{propagate--resample} filter
due to the order of operation of its steps:

\begin{APF*}
\begin{longlist}
\item[\textit{Step 1} (Resample).] $\{{\tilde x}_{t}^{(i)}\}_{i=1}^N$ from
$\{x_{t}^{(i)}\}_{i=1}^N$ with weights
\[
\tilde{w}_{t+1}^{(i)} \propto p\bigl(y_{t+1}|g\bigl(x_{t}^{(i)}\bigr)\bigr).
\]
\item[\textit{Step 2} (Propagate).] $\{{\tilde x}_{t}^{(i)}\}_{i=1}^N$ to $\{
{\tilde x}_{t+1}^{(i)}\}_{i=1}^N$ via\break $p(x_{t+1}|{\tilde x}_{t})$.
\item[\textit{Step 3} (Resample).] $\{{\tilde x}_{t+1}^{(i)}\}_{i=1}^N$
with weights
\[
w_{t+1}^{(i)} \propto\frac{p(y_{t+1}|{\tilde
x}_{t+1}^{(i)})}{p(y_{t+1}|g({\tilde x}_{t}^{(i)}))}.
\]
\end{longlist}
%
\end{APF*}

\subsubsection*{Auxiliary particle filter (APF)}
The APF of Pitt and Shephard (\citeyear{PS1999}) uses a different representation of
the joint filtering distribution of $(x_t,x_{t+1})$ as
\begin{eqnarray*}
&&p(x_t,x_{t+1}|y^{t+1})\\
&&\quad\propto
p(x_{t+1}|x_{t},y^{t+1})p(x_{t}|y^{t+1}) \\
&&\quad= p(x_{t+1}|x_{t},y^{t+1})p(y_{t+1}|x_{t})p(x_{t}|y^{t}).
\end{eqnarray*}
Our view of the APF is as follows: starting with a particle
approximation of $p(x_t|y^t)$, draws from the smoothed distribution of
$p(x_{t}|y^{t+1})$ are obtained by resampling the particles with
weights proportional to the predictive $p(y_{t+1}|x_t)$. These
resampled particles are then propagated forward via
$p(x_{t+1}|x_t,y^{t+1})$. The APF is therefore a \textit{resample--propagate} filter.
Using the terminology of Pitt and Shephard (\citeyear{PS1999}), the above
representation is an optimal, \textit{fully adapted} strategy where exact
samples from $p^N(x_{t+1} | y^{t+1})$ were obtained, avoiding an
importance sampling step. This is possible if both the predictive and
propagation densities were available for evaluation and
sampling.\looseness=1

In general, this is not the case and Pitt and Shephard proposed the use
of an importance function $p(y_{t+1}|{\hat\mu}_{t+1}=g(x_t))$ for the
resampling step
based on a \textit{best guess} for $x_{t+1}$ defined by ${\hat\mu
}_{t+1}=g(x_{t})$. This could be, for example, the expected value, the
median or mode of the state evolution. The resampled particles would
then be propagated with a second proposal defined by
$p(x_{t+1}|x_{t})$, leading to the following algorithm:

Two main ideas make the APF an attractive approach: (i) the
current observation $y_{t+1}$ is used in the proposal of the first
resampling step and (ii) due to the pre-selection in step 1, only
``good'' particles are propagated forward. The importance of this
second point will prove very relevant in the success of our proposed approach.

\subsection{Sequential Parameter Learning Review}\label{sec2.2}
Sequential estimation of fixed parameters $\theta$ is notoriously
difficult. Simply including $\theta$ in the particle set is a natural
but unsuccessful solution, as the absence of a state evolution implies
that we will be left with an ever-decreasing set of atoms in the
particle approximation for $p(\theta|y^t)$. Important developments in
this direction appear in Liu and West (\citeyear{LW2001}), Storvik (\citeyear{S2002}), Fearnhead
(\citeyear{F2002}), Polson, Stroud and
M\"uller (\citeyear{PSM2008}), Johannes and Polson (\citeyear{JP2008}) and Johannes, Polson and
Yae (\citeyear{JPY2008}), to cite a few. We now review two popular alternatives to
learn about $\theta$:

\begin{SF*}
\begin{longlist}
\item[\textit{Step 1} (Propagate).] $\{x_{t}^{(i)}\}_{i=1}^N$ to $\{{\tilde
x}_{t+1}^{(i)}\}_{i=1}^N$ via\break
$q(x_{t+1}|x_{t}^{(i)},\theta^{(i)},y^{t+1})$.
\item[\textit{Step 2} (Resample).] $\{(x_{t+1}, s_{t})^{(i)}\}_{i=1}^N$ from
$\{({\tilde x}_{t+1},\break s_{t})^{(i)}\}_{i=1}^N$ with weights
\[
w_{t+1}^{(i)} \propto\frac{p(y_{t+1}|{\tilde x}_{t+1}^{(i)},\theta
)p({\tilde x}_{t+1}^{(i)}|x_{t}^{(i)},\theta)}{q({\tilde
x}_{t+1}^{(i)}|x_{t}^{(i)},\theta,y^{t+1})}.
\]
\item[\textit{Step 3} (Propagate).] Sufficient statistics
$s_{t+1}^{(i)}=\break\mathcal{S}(s_{t}^{(i)},x_{t+1}^{(i)},y_{t+1})$.
\item[\textit{Step 4} (Sample).] $\theta^{(i)}$ from $p(\theta| s_{t+1}^{(i)})$.
\end{longlist}
\end{SF*}

\subsubsection*{Storvik's filter}
Storvik (\citeyear{S2002}) (similar ideas appear in Fearnhead, \citeyear{F2002}) assumes that
the posterior distribution of $ \theta$ given $ x^t$ and $y^t$ depends
on a low-dimensional set of sufficient statistics that can be
recursively updated. This recursion for
sufficient statistics is defined by $ s_{t+1} = \mathcal{S}( s_t ,
x_{t+1} , y_{t+1})$, leading to the above algorithm.
Notice that the proposal $q(\cdot)$ is conditional on $y_{t+1}$, but
this is still a propagate--resample filter.

\subsubsection*{Liu and West's filter}
Liu and West (\citeyear{LW2001}) suggest a kernel approximation $p(\theta|y^t)$
based on a mixture of multivariate normals. This idea is used in the
context of the APF. Specifically, let $\{(x_t,\theta_t)^{(i)}\}_{i=1}^{N}$
be particle draws from $p(x_t,\theta|y^t)$. Hence, the posterior for
$\theta$ can be approximated by the mixture distribution
\[
p(\theta|y^t) = \sum_{j=1}^N N\bigl(m^{(j)};h^{2}V_{t}\bigr),
\]
where $m^{(j)} = a\theta_{t}^{(j)}+(1-a){\tilde\theta}_{t}$,
${\tilde\theta}_t=\sum_{j=1}^{N} \theta_t^{(j)}/N$\vspace*{-1pt} and $V_{t}=\sum
_{j=1}^{N} (\theta_t^{(j)}-{\bar
\theta}_t)(\theta_t^{(j)}-{\bar\theta}_t)'/N$.
The constants $a$ and $h$ measure, respectively, the extent of the
shrinkage and the degree of overdispersion of the mixture (see Liu and
West, \citeyear{LW2001} for a detailed discussion of the choice of $a$ and $h$).
The idea is to use the mixture approximation to generate fresh samples
from the current posterior in an attempt to avoid particle decay.
The algorithm is summarized in the next page.
The main attraction of Liu and West's filter is its generality, as it
can be implemented in any state-space model. It also takes advantage of
APF's resample--propagate framework and can be considered a benchmark
in the current literature:

\begin{LWF*}
\begin{longlist}
\item[\textit{Step 1} (Resample).] $\{({\tilde x}_{t},{\tilde\theta
}_{t})^{(i)}\}_{i=1}^N$ from $\{(x_{t},\break\theta_{t})^{(i)}\}_{i=1}^N$
with weights
\[
w_{t+1}^{(i)} \propto
p\bigl(y_{t+1}|g\bigl(x_{t}^{(i)}\bigr),m^{(i)}\bigr).
\]
\item[\textit{Step 2} (Propagate).]\mbox{}
\begin{longlist}
\item[(2.1)] $\{{\tilde\theta}^{(i)}_{t}\}_{i=1}^N$ to $\{{\hat\theta
}^{(i)}_{t+1}\}_{i=1}^N$ via $N({\tilde m}^{(i)},V)$;
\item[(2.2)] $\{{\tilde x}_{t}^{(i)}\}_{i=1}^N$ to $\{{\hat
x}_{t+1}^{(i)}\}_{i=1}^N$ via $p(x_{t+1}|{\tilde x}_{t}^{(i)},{\hat
\theta}^{(i)}_{t+1})$.
\end{longlist}
\item[\textit{Step 3} (Resample).] $\{(x_{t+1},\theta_{t+1})^{(i)}\}
_{i=1}^N$ from\break $\{({\hat x}_{t+1},{\hat\theta_{t+1}})^{(i)}\}_{i=1}^N$
with weights
\[
w_{t+1}^{(i)} \propto\frac{p(y_{t+1}|{\hat x}_{t+1}^{(i)},{\hat\theta
}_{t+1}^{(i)})}{p(y_{t+1}|g({\tilde x}_{t}^{(i)}),{\tilde m}^{(i)})}.
\]
\end{longlist}
%
\end{LWF*}

\section{Particle Learning and Smoothing}
\label{sec:general}
Our proposed approach for filtering and learning relies on two main
insights: (i) conditional sufficient statistics are used to represent
the posterior of $\theta$. Whenever possible, sufficient statistics for
the latent states are also introduced, increasing the efficiency of our
algorithm by reducing the variance of sampling weights in what can be
called a Rao--Blackwellized filter.
(ii) We use a resample--propagate framework and attempt to build
perfectly adapted filters whenever possible in trying to obtain exact
samples from our particle approximation when moving from $p^N( x_t ,
\theta| y^t )$ to $p^N( x_{t+1} , \theta| y^{t+1} )$.
This avoids sample importance re-sampling and the associated ``decay''
in the particle approximation.
As with any particle method, there will be accumulation of Monte Carlo
error and this has to be analyzed on a case-by-case basis.
Simply stated, PL builds on the ideas of Johannes and Polson (\citeyear{JP2008}) and
creates a fully adapted extension of the APF to deal with parameter uncertainty.
Without delays, PL can be summarized as follows, with details provided
in the following sections:

\begin{PL*}
\begin{longlist}
\item[\textit{Step 1} (Resample).] $\{{\tilde z}_t^{(i)}\}_{i=1}^N$ from
$z_t^{(i)}=(x_t,s_t,\break\theta)^{(i)}$ with weights
$w_t \propto p(y_{t+1}|z_t^{(i)})$.\vspace*{1pt}
\item[\textit{Step 2} (Propagate).] ${\tilde x}_t^{(i)}$ to $x_{t+1}^{(i)}$
via $p(x_{t+1}|{\tilde z}_t^{(i)},\break y_{t+1})$.
\item[\textit{Step 3} (Propagate).] Sufficient statistics
$s_{t+1}^{(i)} =\break \mathcal{S}({\tilde
s}_t^{(i)},x_{t+1}^{(i)},y_{t+1})$.
\item[\textit{Step 4} (Sample).] $\theta^{(i)}$ from $p(\theta|s_{t+1}^{(i)})$.
\end{longlist}
\end{PL*}

Due to our initial resampling of states and sufficient statistics, we
would end up with a more representative set of propagated
sufficient statistics when sampling parameters than Storvik's filter.

\subsection{Discussion}
Assume that at time $t$, after observing $y^t$, we have a particle
approximation $p^N(z_t| y^t)$, given by $\{z_t^{(i)}\}_{i=1}^N$.
Once $y_{t+1}$ is observed, PL updates the above approximation using
the following resample--propagate rule:
%
\begin{equation} \label{eq:resample}
p(z_t|y^{t+1}) \propto p(y_{t+1}|z_t) p(z_t|y^t)
\end{equation}
and
%
\begin{eqnarray}\label{eq:propagate}
p(z_{t+1}|y^{t+1}) &=& \int p(s_{t+1}|x_{t+1},s_t,y_{t+1})\nonumber\\
&&\hphantom{\int}
{}\cdot p(x_{t+1}|z_t, y_{t+1})\\
&&\hphantom{\int}
{}\cdot p(z_t|y^{t+1}) \,dx_{t+1} \,dz_t.\nonumber
\end{eqnarray}

From (\ref{eq:resample}), we see that an updated approximation
$p^N(z_t| y^{t+1})$ can be obtained by resampling the
current particles set with weights proportional to the predictive
$p(y_{t+1}|z_t)$.
This updated approximation is used in (\ref{eq:propagate}) to generate
propagated samples from the posterior
$p(x_{t+1}|z_t,y_{t+1})$ that are then used to update $s_{t+1}$,
deterministically, by the recursive map $\mathcal{S}(\cdot)$,
which in (\ref{eq:propagate}) we denote by $p( s_{t+1}|x_{t+1}, s_t, y_{t+1})$.
However, since $s_t$ and $x_{t+1}$ are random variables, the
conditional sufficient statistics $s_{t+1}$ are also random and
are replenished, essentially as a state, in the filtering step. This is
the key insight for handling the learning of $\theta$.
The particles for $s_{t+1}$ are sequentially updated with resampled
$s_t$ particles and propagated
and replenished $x_{t+1}$ particles and updated samples from $p(\theta
|s_{t+1})$ can be obtained at the end of the filtering step.

By resampling first we reduce the compounding of approximation errors
as the states are propagated after being ``informed'' by $y_{t+1}$, as
in APF.
To clarify the notion of full-adaptation, we can rewrite the problem of
updating the particles $\{z_t^{(i)}\}_{i=1}^N$ to\break $\{z_{t+1}^{(i)}\}
_{i=1}^N$ as the problem of obtaining samples from the target
$p(x_{t+1},z_t| y^{t+1})$ based on draws from\vspace*{1pt} the proposal $p(z_t |
y^{t+1}) p(x_{t+1}|z_t,y^{t+1}),$ yielding importance weights
%
\begin{equation}
\label{eq:IS}
\quad w_{t+1} \propto\frac{p(x_{t+1},z_t|y^{t+1})}{p(z_t | y^{t+1})
p(x_{t+1}|z_t,y^{t+1})} = 1,
\end{equation}
and therefore, exact draws.
Sampling from the proposal is done in two steps: first draws
$z_t^{(i)}$ from\break $p(z_t | y^{t+1})$ are simply obtained by resampling
the particles $\{z_t^{(i)}\}_{i=1}^N$ with weights proportional
to\break
$p(y_{t+1}|z_t)$; we can then sample $x_{t+1}^{(i)}$ from\break
$p(x_{t+1}|z_t,y^{t+1})$.
Finally, updated samples for $s_{t+1}$ are obtained as a function of
the samples of $x_{t+1}$, with weights $1/N$, which prevents particle
degeneracies in the estimation of $\theta$.
This is a feature of the ``resample--propagate'' mechanism of PL. Any
propagate--resample strategy will lead to decay in the particles of
$x_{t+1}$ with significant negative effects on $p^N(\theta|s_{t+1})$.
This strategy will only be possible whenever both $p(y_{t+1}|z_t)$ and
$p(x_{t+1}|z_t,y^{t+1})$ are analytically tractable, which is the case
in the classes of models considered here.

Convergence properties of the algorithm are\break straightforward to
establish. The choice of particle
size $N$ to achieve a desired level of accuracy depends, however, on
the speed
of Monte Carlo accumulation error. In some cases this will be uniformly
bounded. In others,
a detailed simulation experiment has to be performed. The error will depend
on a number of factors. First, the usual signal-to-noise ratio with the
smaller the value leads
to larger accumulation. Section \ref{sec:CDLM} provides detailed simulation evidence
for the models in question.
Second, a source of Monte Carlo error can appear from using a particle
approximation to
the initial state and parameter distribution. This error is common to
all particle methods.
At its simplest level our algorithm only requires samples $ \theta
^{(i)} $ from the prior
$p(\theta)$. However, a natural class of priors for diffuse situations
are mixtures of the form
$p( \theta) = \int p(\theta| z_0)p(z_0)\, d z_0 $, with the conditional
$p( \theta| z_0 )$ chosen
to be conditionally conjugate. This extra level of analytical
tractability can lead to substantial
improvements in the initial\vspace*{1pt} Monte Carlo error. Particles $z_0^{(i)}$
are drawn from $p( z_0)$ and
then resampled from the predictive and then propagated. Mixtures of
this form are very flexible and
allow for a range of nonconjugate priors. We now turn to specific examples.

\begin{example}[(First order DLM)]\label{ex1}
For illustration, consider first the simple first order dynamic linear
model, also known as the local level model (West and Harrison, \citeyear{WH1997}), where
\begin{eqnarray*}
(y_{t+1} |x_{t+1},\theta) & \sim& N(x_{t+1},\sigma^2), \\
(x_{t+1}|x_t,\theta) & \sim& N(x_{t},\tau^2),
\end{eqnarray*}
with $\theta=(\sigma^2,\tau^2)$,
$x_0 \sim N(m_0,C_0)$,
$\sigma^{2} \sim \operatorname{IG}(a_0,b_0)$
and $\tau^{2} \sim \operatorname{IG}(c_0,d_0)$. The hyperparameters $m_0$, $C_0$,
$a_0$, $b_0$, $c_0$ and $d_0$ are kept fixed and known.
It is straightforward to show that
\begin{eqnarray*}
(y_{t+1}|x_t,\theta) &\sim& N(x_t,\sigma^2+\tau^2) \quad \mbox{and} \\
(x_{t+1}|y_{t+1},x_t,\theta) &\sim& N(\mu_t,\omega^2),
\end{eqnarray*}
where $\mu_t=\omega^2(\sigma^{-2}y_{t+1}+\tau^{-2}x_t)$, $\omega^{-2}=
\sigma^{-2}+\tau^{-2}$.
Also, for scales
\begin{eqnarray*}
(\sigma^2|y^{t+1},x^{t+1}) &\sim& \operatorname{IG}(a_{t+1},b_{t+1}) \quad \mbox{and}
\\
(\tau^2|y^{t+1},x^{t+1}) &\sim& \operatorname{IG}(c_{t+1},d_{t+1}),
\end{eqnarray*}
where $a_{t+1}=a_t + 1/2$, $c_{t+1}=c_t+1/2$,
$b_{t+1}=b_t+0.5(y_{t+1}-x_{t+1})^2$ and
$d_{t+1}=d_t+0.5(x_{t+1}-x_t)^2$. Therefore, the
vector of conditional sufficient statistics $s_{t+1}$ is 5-dimensional
and satisfies the following deterministic recursions:
$s_{t+1}= s_t + (y_{t+1}^2,y_{t+1}x_{t+1},\break x_{t+1}^2,x_t^2,x_{t+1}x_t)$.
Finally, notice that, in both,\break $p(y_{t+1}|x_t)$ and
$p(x_{t+1}|x_t,y^{t+1})$ are available for evaluation and sampling, so
that a fully adapted version of PL can be implemented.
\end{example}

\subsection{State Sufficient Statistics}
A more efficient approach, whenever possible, is to marginalize states
and just track conditional state sufficient statistics.
In the pure filtering case, Chen and Liu (\citeyear{CL2000}) use a similar approach.
Here we use the fact that
\[
p(x_{t}|y^{t})=\int p(x_{t}|s_{t}^{x})p(s_{t}^{x}|y^{t})\,ds_{t}^{x}.
\]
Thus, we are interested in the distribution $p( s_{t}^{x}|y^{t})$.
The filtering recursions are given by
\begin{eqnarray*}
p( s_{t+1}^{x}|y^{t+1}) &=& \int p(
s_{t+1}^{x}|s_{t}^{x},x_{t+1},y_{t+1})\\
&&\hphantom{\int}{}\cdot p( s_{t}^{x},x_{t+1}|y^{t+1})
\,ds_{t}^{x}\,dx_{t+1}.
\end{eqnarray*}
We can decompose $p(s_t^x,x_{t+1}|y^{t+1})$ as proportional to
\[
p(y_{t+1}|s_{t}^{x}) p( x_{t+1}|s_{t}^{x},y_{t+1}) p(s_{t}^{x}|y^{t}),
\]
where we have an extra level of marginalization.
Instead of marginalizing $x_{t}$, you now marginalize over $s_{t}^{x}$
and $x_{t+1}$.
For this to be effective, we need the following conditional posterior:
\[
p( x_{t+1}|s_{t}^{x},y_{t+1}) =\int p( x_{t+1}
|x_{t},y_{t+1}) p( x_{t}|s_{t}^{x}) \,dx_{t}.
\]
We can then proceed with the particle learning algorithm.
Due to this Rao--Blackwellization step, the weights are flatter in the
first stage, that is, $p( y_{t+1}|s_{t}^{x})$ versus $p(
y_{t+1}|x_{t})$ increasing the efficiency of the algorithm.

\setcounter{example}{0}
\begin{example}[(Cont.)]
Recalling $(x_t|\theta) \sim N(m_t,\break C_t)$, then it is straightforward to
see that
$(y_{t+1}|m_t,\break C_t,\theta) \sim N(m_t,C_t+\sigma^2+\tau^2)$, so
$s_t^x=(m_t,C_t)$.
The recursions for the state sufficient statistics vector $s_t^x$ are
the well-known Kalman recursions, that is, $m_{t+1}=(1-A_{t+1})m_t +
A_{t+1}y_{t+1}$ and
$C_{t+1}=A_{t+1}\sigma^2$, where $A_{t+1}=(C_t+\tau^2)/(C_t+\tau
^2+\sigma^2)$ is the Kalman gain.
\end{example}

\subsection{Smoothing}
\label{sec:smooth}
Smoothing, that is, estimating the states and parameters conditional on
all available information, is characterized by
$p(x^T,\theta|y^T)$, with $T$ denoting the last observation.

After one sequential pass through the data, our particle approximation
computes samples from\break $p^{N} ( x_{t} , s_{t} |
y^{t} ) $ for all $t \leq T$. However, in many situations, we are
required to obtain full smoothing\vspace*{1pt}
distributions $p( x^T| y^T ) $ which are typically carried out by a
MCMC scheme.
We now show that our filtering strategy provides a direct backward
sequential pass to sample from the target smoothing distribution. To
compute the marginal smoothing distribution, we write the joint
posterior of $(x^T,\theta)$ as
\[
p(x^T,\theta|y^T) =
\prod_{t=1}^{T-1}{p(x_t|x_{t+1},\theta,y^{t})}p(x_T,\theta|y^T ) .
\]
By Bayes' rule and conditional independence, we have
\[
p(x_t|x_{t+1},\theta,y^{t}) \propto p(x_{t+1}|x_t, \theta, y^{t})
p(x_t|\theta, y^{t} ).
\]
We can now derive a recursive backward sampling algorithm to jointly
sample from $p( x^T,\theta|y^T)$ by sequentially sampling from
filtered particles with\break weights proportional to $p(x_{t+1}|x_t,\theta,
y^t)$. In detail,
randomly choose, at time $T$, $(\tilde{x}_T,\tilde{s}_T)$ from the
particle approximation $p^{N} ( x_{T} , s_{T} |
y^{T} )$ and sample $\tilde{\theta} \sim p(\theta| \tilde{s}_T)$.
Then, for $t=T-1,\dots,1$, choose $\tilde{x}_t = x_t^{(i)}$ from the
filtered particles $\{x_t^{(i)},i=1,\ldots,N\}$ with weights
$w_{t|t+1}^{(i)} \propto p(\tilde{x}_{t+1}|x_t^{(i)},\tilde{\theta})$:

\begin{PS*}
\begin{longlist}
\item[\textit{Step 1} (Forward filtering).]
Sample $\{(x^T,\theta)^{(i)}\}_{i=1}^N $\break via \textit{particle
learning}.
\item[\textit{Step 2} (Backwards smoothing).]
For each pair $(x_T,\break\theta)^{(i)}$ and $t=T-1,\ldots,1$,
resample $x_t^{(i)}$ from $\{x_t^{(j)}\}_{j=1}^N$ with weights
\[
w_{t|t+1}^{(j)} \propto p\bigl(x_{t+1}^{(i)}|x_t^{(j)},\theta^{(i)}\bigr).
\]
\end{longlist}
\end{PS*}

This algorithm is an extension of Godsill, Doucet and West (\citeyear{GDW2004}) to
state space models where the fixed parameters are unknown. See also
Briers, Doucet and Maskell (\citeyear{BDM2010}) for an alternative SMC smoother. Both
SMC smoothers are $O(TN^2)$, so
the computational time to obtain draws from $p(x^T|y^T)$
is expected to be much larger than the computational time to obtain
draws from
$p(x_t|y^t)$, for $t=1,\ldots,T$, from standard SMC filters.
An $O(TN)$ smoothing algorithm has recently been introduced by
Fearnhead, Wyncoll and Tawn (\citeyear{FWT2008}).

\setcounter{example}{0}
\begin{example}[(Cont.)]
For $t=T-1,\ldots,2,1$, it is easy to see that
$(x_t|x_{t+1},y^T,\theta) \sim N(a_t,D_t\tau^2)$\vspace*{1.5pt}
and $(x_t|y^T,\theta) \sim N(m_t^T,C_t^T)$, where
$a_t=(1-D_t)m_t+D_t x_{t+1}$
$m_t^T=(1-D_t)m_t+D_t m_{t+1}^T$,
$C_t^T=(1-D_t)C_t+D_t^2C_{t+1}^T$,
and $D_t=C_t/(C_t+\tau^2)$.\vspace*{1pt} Finally, $m_T^T=m_T$ and $C_T^T=C_T$.
\end{example}

\subsection{Model Monitoring}
\label{sec:monitoring}

The output of PL can be used for sequential predictive problems but is
also key in the computation of Bayes factors for model assessment in
state space models. Specifically, the marginal predictive for a given
model $\mathcal{M}$ can be approximated via
\[
p^N ( y_{t+1} | y^t,\mathcal{M}) = \frac{1}{N} \sum_{i=1}^N p\bigl( y_{t+1} | (
x_t , \theta)^{(i)},\mathcal{M}\bigr).
\]
This then allows the computation of a SMC approximation to the Bayes
factor $B_{t+1}$
or sequential likelihood ratios for competing models $\mathcal{M}_0$ and
$\mathcal{M}_1$ (see, e.g., West, \citeyear{W1986}):
\[
B_{t+1}= \frac{p( y_1,\ldots,y_{t+1}|\mathcal{M}_1 )}{ p(y_1,\ldots
,y_{t+1}|\mathcal{M}_0 )},
\]
where $p(y_1,\ldots,y_{t+1}|\mathcal{M}_i ) = \prod_{j=1}^{t+1} p( y_j |
y^{j-1},\mathcal{M}_i)$, for either model.

\begin{MM*}
\begin{longlist}
\item[\textit{Step 1}.] Compute the predictive using
\[
p^N ( y_{t+1} | y^t ) = \frac{1}{N} \sum_{i=1}^N p\bigl( y_{t+1} | ( x_t ,
\theta)^{(i)} \bigr).
\]
\item[\textit{Step 2}.] Compute the marginal likelihood
\[
p^N(y_1,\ldots,y_{t+1})=\prod_{j=1}^{t+1} p^N(y_{j+1}|y^j).
\]
\end{longlist}
\end{MM*}

An important advantage of PL over MCMC\break schemes is that it directly
provides the filtered joint posteriors $p(x_t,\theta| y^t)$
and, hence, $ p( y_{t+1} | y^t ) $, whereas MCMC would have to be
repeated $T$ times to make that available.

\section{Conditional Dynamic Linear Models}
\label{sec:CDLM}
We now explicitly derive our PL algorithm in a class of conditional
dynamic linear models which are an extension of the models
considered in West and Harrison (\citeyear{WH1997}). This consists of a vast class
of models that embeds many of the commonly used dynamic models.
MCMC via Forward-filtering\break Backward-sampling (Carter and Kohn,
\citeyear{CK1994};\break
Fr\"uhwirth-Schnatter, \citeyear{FS1994}) or mixture Kalman filtering (MKF) (Chen
and Liu, \citeyear{CL2000}) are the current methods of use for the estimation of
these models.
As an approach for filtering, PL has a number of advantages. First, our
algorithm is
more efficient, as it is a perfectly-adapted filter. Second, we extend
MKF by including learning about fixed parameters and smoothing for states.

The conditional DLM defined by the observation and
evolution equations takes the form of a linear system conditional on an
auxiliary state $\lambda_{t+1}$,
\begin{eqnarray*}
(y_{t+1}|x_{t+1},\lambda_{t+1},\theta) & \sim& N(F_{ \lambda_{t+1} }
x_{t+1}, V_{ \lambda_{t+1} }), \\
(x_{t+1}|x_t,\lambda_{t+1},\theta) & \sim& N(G_{ \lambda_{t+1} }
x_{t},W_{ \lambda_{t+1} }),
\end{eqnarray*}
with $\theta$ containing $F$'s, $G$'s, $V$'s and $W$'s.
The\break marginal distribution of observation error and state shock
distribution are any combination of
normal, scale mixture of normals or discrete mixture of normals
depending on
the specification of the distribution on the auxiliary state variable $
p ( \lambda_{t+1} | \theta) $, so that
\begin{eqnarray*}
p(y_{t+1}|x_{t+1},\theta) &=& \int f_N(y_{t+1};F_{ \lambda_{t+1} }
x_{t+1}, V_{ \lambda_{t+1} })\\
&&\hphantom{\int}{}\cdot p( \lambda_{t+1} | \theta)\,d\lambda_{t+1}.
\end{eqnarray*}
Extensions to hidden Markov specifications where
$ \lambda_{t+1}$ evolves according to $p ( \lambda_{t+1} | \lambda_{t},
\theta) $ are straightforward and are discussed in Example \ref{ex:dfm} below.

\subsection{Particle Learning in CDLM}
In CDLMs the state filtering and parameter learning problem is
equivalent to a filtering problem
for the joint distribution of their respective sufficient statistics.
This is a direct result of the factorization of the full joint
\[
p( x_{t+1} , \theta, \lambda_{t+1}, s_{t+1} ,s^x_{t+1} | y^{t+1})
\]
as a sequence of conditional distributions
\[
p( \theta| s_{t+1} ) p( x_{t+1} | s^x_{t+1}, \lambda_{t+1} ) p( \lambda
_{t+1},s_{t+1} ,s^x_{t+1} | y^{t+1}).
\]
Here the conditional sufficient statistics for states ($s_t^x$) and
parameters ($s_t$) satisfy deterministic updating rules
%
\begin{eqnarray}
s_{t+1}^x&=&\mathcal{K}( s_{t}^x,\theta,\lambda_{t+1},y_{t+1}
), \label{eq:KF} \\
s_{t+1}&=&\mathcal{S}( s_{t},x_{t+1},\lambda_{t+1},y_{t+1}),
\label{eq:CSS}
\end{eqnarray}
where $ \mathcal{K}( \cdot)$ denotes the Kalman filter recursions
and $ \mathcal{S} ( \cdot) $ our recursive update of the sufficient statistics.
More specifically, define $s_{t}^{x} = ( m_{t},C_{t}) $ as
Kalman filter first
and second moments at time $t$. Conditional on $\theta$, we then have
\[
( x_{t+1} | s^x_{t+1},\lambda_{t+1},\theta,) \sim N( a_{t+1} , R_{t+1}),
\]
where
$a_{t+1} =G_{\lambda_{t+1}} m_t$ and\vspace*{1.5pt} $R_{t+1} = G_{\lambda_{t+1} } C_t
G^\prime_{\lambda_{t+1} } +W_{ \lambda_{t+1} } $.
Updating state sufficient statistics $ ( m_{t+1} , C_{t+1} )$ is
achieved by
%
\begin{eqnarray}
m_{t+1} &=&G_{ \lambda_{t+1} } m_t + A_{t+1} ( y_{t+1} - e_t
), \label{eq:Kalman1}\\
C_{t+1}^{-1} &=& R_{t+1}^{-1} + F_{ \lambda_{t+1} }^\prime F_{ \lambda
_{t+1} } V_{ \lambda_{t+1} }^{-1} \label{eq:Kalman2},
\end{eqnarray}
with Kalman gain matrix $A_{t+1} = R_{t+1} F_{ \lambda_{t+1} }
Q_{t+1}^{-1}$, $e_t=F_{ \lambda_{t+1} } G_{ \lambda_{t+1} } m_t$, and
$Q_{t+1} = F_{\lambda_{t+1}} R_{t+1} F_{\lambda_{t+1}} +\break V_{\lambda_{t+1}}$.

We are now ready to define the PL scheme for the CDLMs.
First, assume that the auxiliary state variable is discrete with $
\lambda_{t+1} \sim p( \lambda_{t+1} | \lambda_t, \theta) $. We start,
at time $t$, with a particle approximation for the joint posterior of
$(x_{t}, \lambda_{t}, s_{t} ,s^x_{t},\theta| y^{t})$.
Then we propagate to $t+1$ by first resampling the current particles
with weights proportional to the predictive $p( y_{t+1} | ( \theta,
s^x_t )) $.
This provides a particle approximation to $p( x_t,\theta,\lambda_t,s_t
, s^x_t | y^{t+1} )$, the smoothing distribution.
New states $\lambda_{t+1}$ and $x_{t+1}$ are then propagated through
the conditional posterior distributions
$p( \lambda_{t+1} | \lambda_t , \theta, y_{t+1}) $ and $p( x_{t+1} |
\lambda_{t+1} , x_t , \theta, y_{t+1})$.
Finally, the conditional sufficient statistics are updated according to
(\ref{eq:KF}) and (\ref{eq:CSS}) and new samples for $\theta$
are obtained from $p(\theta|s_{t+1})$. Notice that in the conditional
dynamic linear models all the above densities are available for
evaluation and sampling. For instance, the predictive is computed via
\begin{eqnarray*}
p\bigl( y_{t+1} | ( \lambda_t,s_{t}^{x} ,\theta)^{(i)}\bigr)& =& \sum_{ \lambda_{t+1} } p\bigl( y_{t+1} | \lambda_{t+1} , ( s_{t}^{x} ,
\theta)^{(i)}\bigr)\\
&&\hphantom{\sum_{ \lambda_{t+1} }}{}\cdot p( \lambda_{t+1}|\lambda_t,\theta),
\end{eqnarray*}
where the inner predictive distribution is given by
\begin{eqnarray*}
p( y_{t+1} | \lambda_{t+1} , s_{t}^{x} , \theta )
&=& \int p( y_{t+1} | x_{t+1} , \lambda_{t+1} , \theta )\\
&&\hphantom{\int }{}\cdot p( x_{t+1}| s^x_t , \theta) \,dx_{t+1}.
\end{eqnarray*}

Starting with particle set $\{(x_0,\theta, \lambda
_0,s_0,s_0^{x})^{(i)},i=1,\ldots,N\}$ at time $t=0$,
the above discussion can be summarized in the PL Algorithm \ref{alg1}.
In the general case where the auxiliary state variable $ \lambda_t $ is
continuous, it might not be possible to integrate out $\lambda_{t+1}$
form the predictive in step 1. We extend the above scheme by adding to
the current particle set a propagated particle $ \lambda_{t+1} \sim p(
\lambda_{t+1}|( \lambda_{t}, \theta) ^{(i)}) $ and define the PL
Algorithm~\ref{alg2}.

Both algorithms can be combined with the backward propagation scheme of
Section \ref{sec:smooth} to provide a full draw from the marginal
posterior distribution for all the states given the data, namely, the
smoothing distribution $p(x_1,\ldots,x_T| y^T) $.

\begin{alg}[(CDLM)]\label{alg1}
\begin{longlist}
\item[\textit{Step 1} (Resample).] ${\tilde z}_t^{(i)}$ from
$z_t^{(i)}=(\lambda_t,s_{t}^{x} , \theta)^{(i)}$\break
with weights
\[
w_{t+1}^{(i)} \propto p\bigl( y_{t+1} | ( \lambda_t,s_{t}^{x} ,
\theta)^{(i)}\bigr).
\]
\item[\textit{Step 2} (Propagate).] States
\begin{eqnarray*}
\lambda_{t+1}^{(i)} & \sim& p\bigl(\lambda_{t+1}|({\tilde\lambda}_t,{\tilde
\theta}) ^{(i)} , y_{t+1}\bigr), \\
x_{t+1}^{(i)} &\sim& p\bigl(x_{t+1}|({\tilde x}_t,{\tilde\theta})^{(i)},
\lambda_{t+1}^{(i)},y_{t+1}\bigr).
\end{eqnarray*}

\item[\textit{Step 3} (Propagate).] Sufficient statistics
\begin{eqnarray*}
s_{t+1}^{x(i)} & =&\mathcal{K}\bigl({\tilde s}_{t}^{x(i)},\lambda
_{t+1}^{(i)},{\tilde\theta}^{(i)}, y_{t+1}\bigr), \\
s_{t+1}^{(i)} & =&\mathcal{S}\bigl({\tilde s}_t^{(i)}, x_{t+1}^{(i)},\lambda
_{t+1}^{(i)},{\tilde\theta}^{(i)}, y_{t+1}\bigr).
\end{eqnarray*}

\item[\textit{Step 4} (Propagate).] Parameters $\theta^{(i)} \sim p(\theta
|s_{t+1}^{(i)})$.
\end{longlist}
\end{alg}

\begin{example}[(Dynamic factor model with time-varying loadings)]\label{ex:dfm}
Consider data $y_{t}=(y_{t1},y_{t2})'$, $t=1,\ldots,T$, following a
dynamic factor model with time-varying loadings driven by a
discrete latent state $\lambda_t$ with possible values $\{1,2\}$.
Specifically, we have
\begin{eqnarray*}
(y_{t+1}|x_{t+1},\lambda_{t+1},\theta) & \sim &N(\beta
_{t+1}x_{t+1},\sigma^2I_2),\\
(x_{t+1}|x_t,\lambda_{t+1},\theta) & \sim& N(x_{t}, \sigma_{x}^2),
\end{eqnarray*}
with time-varying loadings ${\beta}_{t+1}=(1,\beta_{\lambda_{t+1}})'$ and
initial state distribution $x_{0} \sim N(m_{0},C_{0})$.
The jumps in the factor loadings are driven by a Markov switching
process $( \lambda_{t+1}|\lambda_{t},\theta)$,
whose transition matrix $\Pi$ has diagonal elements $\operatorname{Pr}(\lambda
_{t+1}=1|\lambda_t=1,\theta)=p$ and
$\operatorname{Pr}(\lambda_{t+1}=2|\lambda_t=2,\theta)=q$.
The parameters are $\theta=(\beta_{1},\beta_{2},\sigma^{2},\tau
^2,p,q)^{\prime}$.
See Carvalho and Lopes (\citeyear{CL2007}) for related Markov switching models.

We are able to marginalize over both
$(x_{t+1} , \lambda_{t+1})$ by using state sufficient statistics
$s_t^x=( m_{t} ,
C_{t}) $ as particles. From the Kalman filter recursions we know that $p(
x_{t}| \lambda^{t}, \theta, y^{t}) \sim N(m_{t},C_{t})$.
The mapping for state sufficient statistics $(m_{t+1},C_{t+1}) =
\mathcal{K} ( m_{t} , C_{t},\lambda_{t+1},\theta,\break y_{t+1} ) $ is
given by the one-step Kalman update as in (\ref{eq:Kalman1}) and (\ref{eq:Kalman2}).
The prior distributions are conditionally conjugate where\vspace*{1pt}
$(\beta_{i}|\sigma^{2} ) \sim N( b_{i0},\sigma^{2} B_{i0})$ for\break $i=1,2$,
$\sigma^{2} \sim \operatorname{IG}(\nu_{00}/2,d_{00}/2)$
and $\tau^2 \sim \operatorname{IG}(\nu_{10}/2,\break d_{10}/2)$.
For the transition probabilities, we assume that $p \sim \operatorname{Beta}(p_{1},p_{2})$
and $q \sim \operatorname{Beta}(q_{1},q_{2})$. Assume that, at time $t$, we have
particles\vspace*{1pt} $\{(x_t,\theta,\lambda_t,s^x_t,s_t)^{(i)}\}_{i=1}^N$,
for $i=1,\ldots,N$, approximating $p(x_t,\theta,\lambda
_t,s_t^x,s_t|y^t)$. The PL algorithm can be described through the
following steps:
\begin{enumerate}
\item\textit{Resampling}: Draw an index $k^i \sim
\operatorname{Mult}(w_t^{(1)},\ldots,\break w_t^{(N)}) $ with weights
$w_t^{(i)} \propto p(y_{t+1}| ( m_{t}, C_{t} , \lambda_{t} ,\break  \theta
)^{(k^i)})$ where
\begin{eqnarray*}
&&p(y_{t+1}|s_t^x ,\lambda_{t},\theta)\\
&&\quad  = \sum_{\lambda_{t+1}=1}^2
f_N(y_{t+1};a,b) p(\lambda_{t+1} |\lambda_t,\theta),
\end{eqnarray*}
where $ f_N(x;a,b)$ denotes the density of the normal distribution with
mean $a$ and variance $b$ and evaluation at the point $x$. Here $a=\beta
_{t+1}m_{t}$ and $b=( C_{t}+\tau^2)\beta_{t+1}\beta_{t+1}' +\sigma^{2} I_{2}$.

\item\textit{Propagating state $\lambda$}:
Draw $\lambda_{t+1}^{(i)}$ from\vspace*{-1pt} $p(\lambda_{t+1}|( s^x_t ,\break \lambda
_{t},\theta)^{(k^i)} ,y_{t+1})$:
\begin{eqnarray*}
\hspace*{-2pt}&&p(\lambda_{t+1}|s^x_t , \lambda_{t} ,
\theta,y_{t+1} ) \\
\hspace*{-2pt}&&\quad \propto f_N\bigl(y_{t+1};\beta_{t+1}m_{t},
( C_{t}+\tau^2) \beta_{t+1}\beta_{t+1}'+\sigma^{2} I_{2}\bigr)\\
\hspace*{-2pt}&&\qquad {}\cdot p(\lambda_{t+1}|\lambda
_t,\theta).
\end{eqnarray*}

\item\textit{Propagating state $x$}:
Draw $x_{t+1}^{(i)}$ from\vspace*{-2pt} $p(x_{t+1}|\lambda_{t+1}^{(i)},\break( s_t^x ,\theta
)^{(k^i)} ,y_{t+1})$.

\item\textit{Propagating sufficient statistics for states}:
The Kalman filter recursions yield
\begin{eqnarray*}
m_{t+1} & = & m_{t} + A_{t+1}(y_{t+1}-\beta_{t+1} m_{t}),\\
C_{t+1} & = & C_{t}+\tau^2 - A_{t+1}Q_{t+1}^{-1} A_{t+1}^{\prime},
\end{eqnarray*}
where
$Q_{t+1}=( C_{t}+\tau^2)\beta_{t+1}\beta_{t+1} +\sigma^{2} I_{2}$
and\break $A_{t+1}=( C_{t}+\tau^2)Q_{t+1}^{-1}\beta_{t+1}$.

\item\textit{Propagating sufficient statistics for parameters}:
The conditional posterior $p( \theta| s_{t} )$, for $i=1,2$, is
decomposed into
\begin{eqnarray*}
p(\beta_{i}| \sigma^{2} , s_{t+1} ) & \sim& N( b_{i,t+1},\sigma
^{2}B_{i,t+1}),\\
p(\sigma^{2}| s_{t+1}) & \sim& \operatorname{IG}(\nu_{0,t+1}/2,d_{0,t+1}/2t), \\
p(\tau^2| s_{t+1}) &\sim& \operatorname{IG}(\nu_{1,t+1}/2,d_{1,t+1}/2), \\
p( p| s_{t+1} ) & \sim& \operatorname{Beta}(p_{1,t+1},p_{2,t+1}), \\
p( q| s_{t+1}) & \sim& \operatorname{Beta}(q_{1,t+1},q_{2,t+1}),
\end{eqnarray*}
with
$B_{i, t+1}^{-1}=B_{it}^{-1}+ x_{t+1}^{2} \mathbb{I}_{\lambda
_{t+1}=i}$, $b_{i,t+1}=B_{i,t+1}\cdot ( B_{it}^{-1} b_{it} +
x_{t}y_{t2} \mathbb{I}_{\lambda_{t+1}=i} )$ and
$\nu_{i,t+1}=\nu_{i,t} + 1$, for $i=1,2$,
$d_{1,t+1}=d_{1t} + (x_{t+1}-x_t)^2$,
$p_{1,t+1} = p_{1t} + \mathbb{I}_{ \lambda_{t}=1,\lambda_{t+1}=1 }$,
$p_{2,t+1} = p_{2t} + \mathbb{I}_{ \lambda_{t}=1,\lambda_{t+1}=2}$,
$q_{1,t+1} = q_{1t} + \mathbb{I}_{ \lambda_{t}=2,\lambda_{t+1}=2}$
$q_{2,t+1} = q_{2t} + \mathbb{I}_{ \lambda_{t}=2,\lambda_{t+1}=1}$ and
$d_{0,t+1}=d_{0t} + \sum_{j=1}^2 [(y_{t+1,2} - b_{j,t+1} x_{t+1})
y_{t+1,2} +$\break $b_{j,t+1} B^{-1}_{j0} + (y_{t+1,1} - x_{t+1})^2] \mathbb
{I}_{\lambda_{t+1}=j}$.
\end{enumerate}
\end{example}

\begin{figure*}

\includegraphics{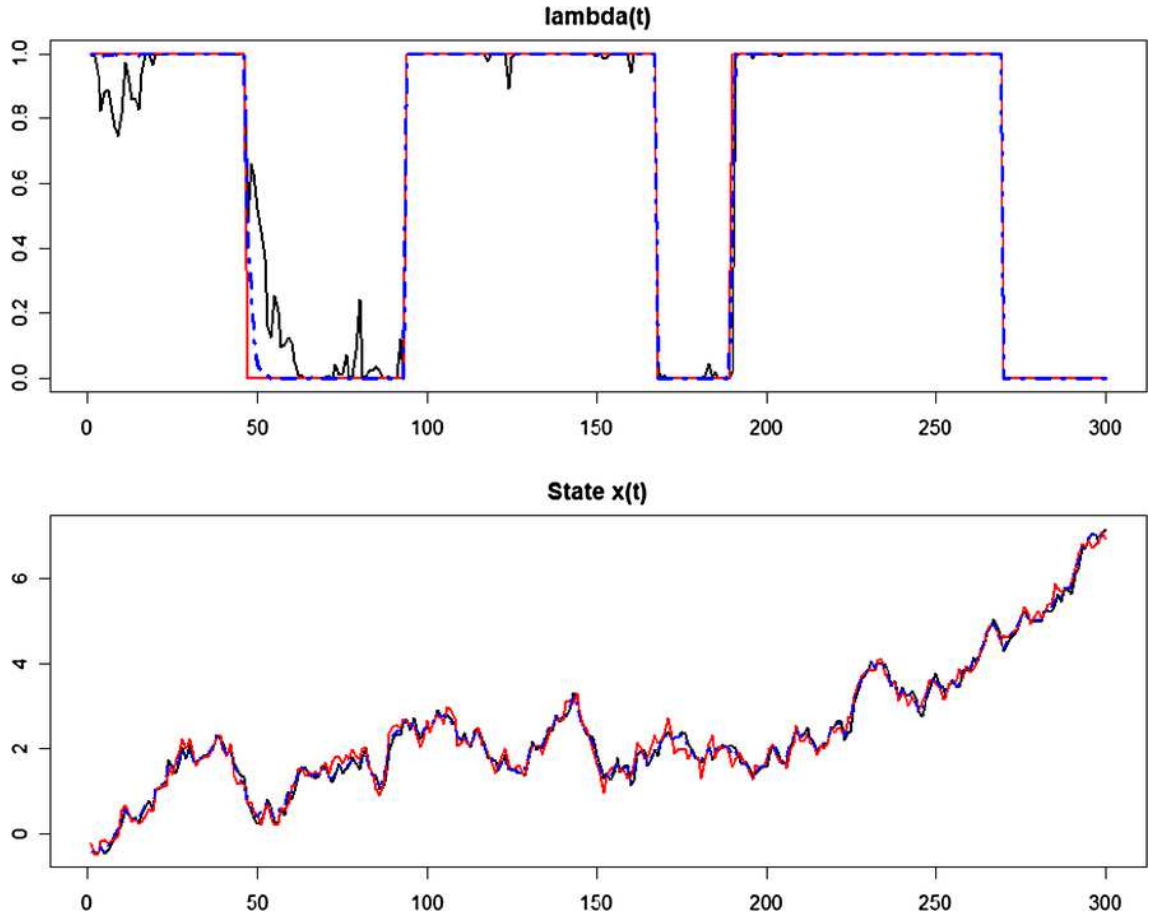}

\caption{Dynamic factor model (state learning).
Top panel: True value of $\lambda_t$ (red line),
$\operatorname{Pr}(\lambda_t=1|y^t)$ (black line) and
$\operatorname{Pr}(\lambda_t=1|y^T)$ (blue line).
Bottom panel: True value of $x_t$ (red line), $E(x_t|y^t)$ (black line)
and $E(x_t|y^T)$ (blue line).}\vspace*{5pt}
\label{fig:fig1}
\end{figure*}

\begin{figure*}[t]

\includegraphics{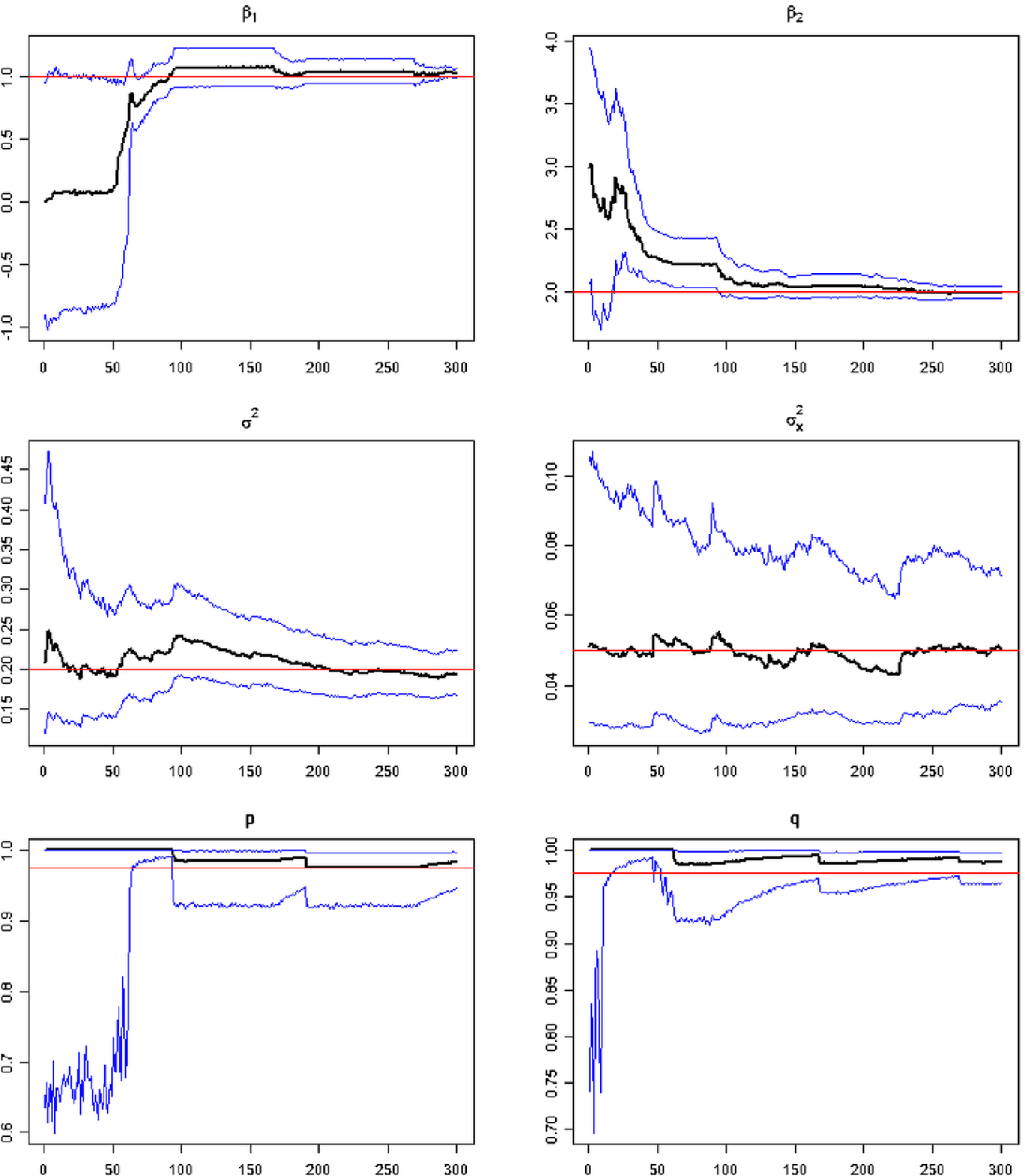}

\caption{Dynamic factor model (parameter learning).
Sequential posterior median (black line) and posterior 95\% credibility intervals (blue lines)
for model parameters $\beta_1$, $\beta_2$, $\sigma^2$, $\tau^2$, $p$ and $q$.
True values are the red lines.}
\label{fig:fig2}
\end{figure*}

Figures \ref{fig:fig1} and \ref{fig:fig2} illustrate the performance of
the PL algorithm.
The first panel of Figure \ref{fig:fig1} displays the true underlying
$\lambda$ process along with filtered and smoothed estimates, whereas
the second panel presents the same information for the common factor.
Figure \ref{fig:fig2} provides the sequential parameter learning plots.

\begin{alg}[(Auxiliary state CDLM)]\label{alg2}
Let $z_t=(\lambda_{t+1},x_t,s_{t}^{x},\theta)$.
\begin{longlist}
\item[\textit{Step 0} (Propagate).] $\lambda_t^{(i)}$ to $\lambda
_{t+1}^{(i)}$ via
$\lambda_{t+1}^{(i)} \sim p( \lambda_{t+1}|\break( \lambda_{t}, \theta)
^{(i)})$.

\item[\textit{Step 1} (Resample).] ${\tilde z}_t^{(i)}$ from $z_t^{(i)}$
with weights\break
$w_{t+1}^{(i)} \propto p(y_{t+1} | {\tilde z}_t^{(i)})$.

\item[\textit{Step 2} (Propagate).] ${\tilde x}_t^{(i)}$ to $x_{t+1}^{(i)}$ via
$p( x_{t+1}|{\tilde z}_t^{(i)},\break y_{t+1})$.

\item[\textit{Step 3} (Propagate).] Sufficient statistics as in PL.
\item[\textit{Step 4} (Propagate).] Parameters as in PL.
\end{longlist}
\end{alg}

\section{Nonlinear Filtering and Learning}
\label{sec:nonlin}
We now extend our PL filter to a general class of nonlinear state space
models, namely, the conditional Gaussian dynamic model (CGDM).
This class generalizes conditional dynamic linear models by allowing
nonlinear evolution equations.
In this context we take advantage of most efficiency gains of PL, as we
are still able to follow the resample/propagate
logic and filter sufficient statistics for $\theta$.
Consider a conditional Gaussian state space model with nonlinear
evolution equation,
\begin{eqnarray}
\qquad (y_{t+1}|x_{t+1},\lambda_{t+1},\theta) & \sim& N(F_{\lambda_{t+1}}
x_{t+1},V_{\lambda_{t+1}}), \\
(x_{t+1}|x_t,\lambda_{t+1},\theta) & \sim &N(G_{\lambda_{t+1}} h(
x_{t}),W_{\lambda_{t+1}}),
\end{eqnarray}
where $h(\cdot)$ is a given nonlinear function and, again, $\theta$
contains $F$'s, $G$'s, $V$'s and $W$'s.
Due to the nonlinearity in the evolution, we are no longer able to work
with state sufficient statistics $s_t^x$, but we are still able to
evaluate the predictive $p(y_{t+1}|x_t,\lambda_t,\theta)$.
In general, take as the particle set the following:
$\{ ( x_t,\theta, \lambda_t, s_t )^{(i)}, i=1,\ldots,N\}$.
For discrete $\lambda$ we can define the following
algorithm:

\begin{figure*}[b]

\includegraphics{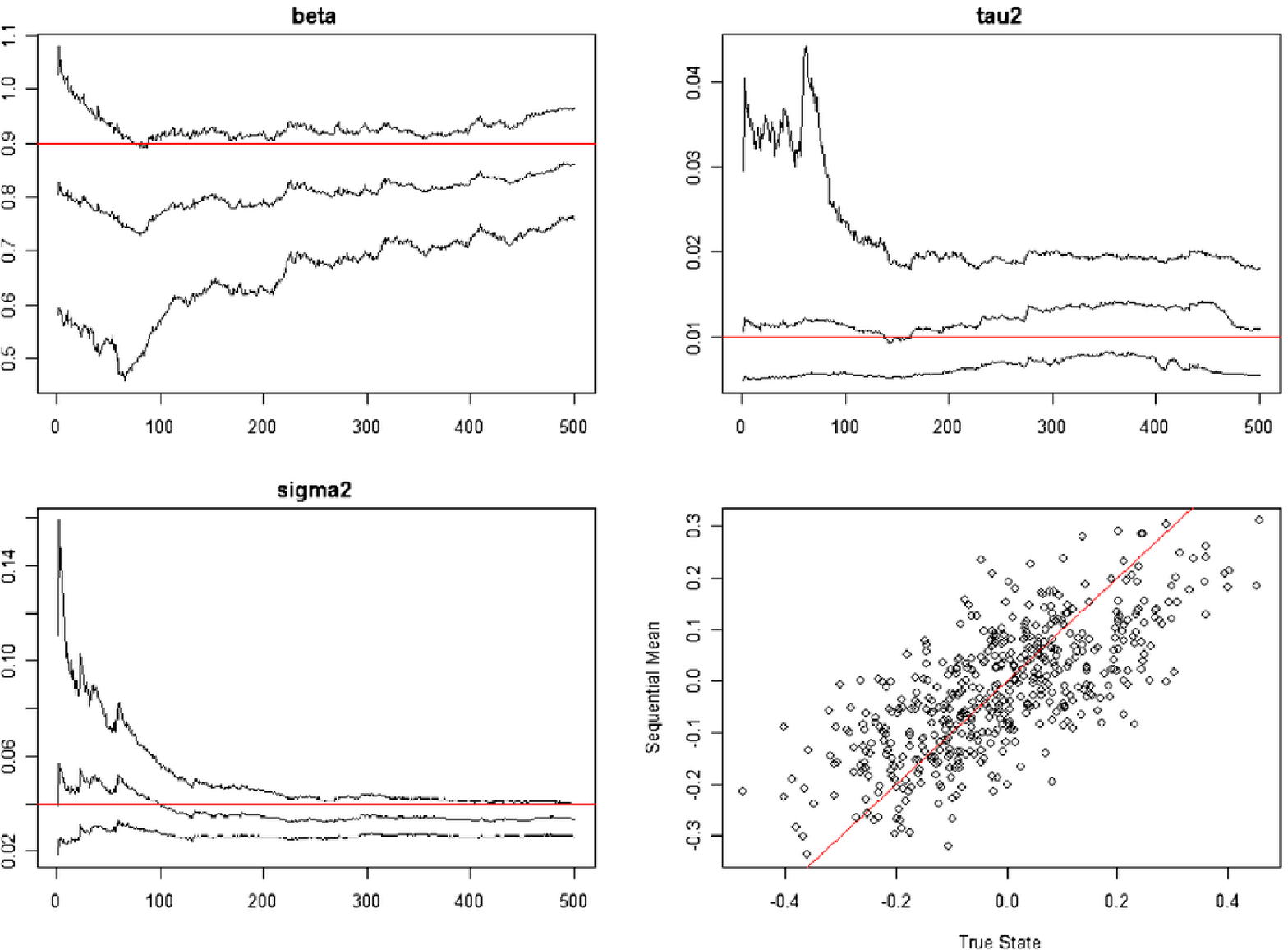}

\caption{Heavy-tailed non-Gaussian, nonlinear model.
Sequential posterior median and posterior 95\% credibility intervals (black lines)
for model parameters $\beta$, $\sigma^2$ and $\tau^2$.  True values are the red lines.
The bottom right panel is the true value of $x_t$ against $E(x_t|y^t)$.}
\label{fig:fig3}
\end{figure*}

\begin{alg}[(CGDM)]\label{alg3}
\begin{longlist}
\item[\textit{Step 1} (Resample).] ${\tilde z}_t^{(i)}$ from $z_t^{(i)}=(
x_{t} ,\lambda_t, \theta)^{(i)}$\break
with weights
\[
w_t^{(i)} \propto p\bigl( y_{t+1} | ( x_{t} ,\lambda_t,
\theta)^{(i)}\bigr).
\]
\item[\textit{Step 2} (Propagate).] States
\begin{eqnarray*}
\lambda_{t+1}^{(i)} & \sim& p\bigl(\lambda_{t+1}|({\tilde\lambda
}_{t},{\tilde\theta})^{(i)},y_{t+1}\bigr), \\
x_{t+1}^{(i)} & \sim& p\bigl( x_{t+1}|( {\tilde x}_{t},{\tilde\theta})
^{(i)},\lambda_{t+1}^{(i)},y_{t+1}\bigr).
\end{eqnarray*}
\item[\textit{Step 3} (Propagate).] Parameter\vspace*{1pt} sufficient statistics as in
Algorithm \ref{alg1}.
\item[\textit{Step 4} (Propagate).] Parameters as in PL.
\end{longlist}
\end{alg}

When  $\lambda$ is continuous, propagate $\lambda_{t+1}^{(i)}$ from $p(
\lambda_{t+1}|\break ( \lambda_{t}, \theta)^{(i)})$, for $i=1,\ldots,N$,
then we  resample the particle $ ( x_t, \lambda_{t+1}, \theta,s_{t} )
^{(i)} $ with the appropriate predictive distribution $ p( y_{t+1} | (
x_{t} , \lambda_{t+1},\theta)^{(i)} ) $ as in Algorithm \ref{alg2}.
Finally, it is straightforward to extend the backward smoothing
strategy of Section \ref{sec:smooth} to obtain\vspace*{1.5pt} samples from $p(x^T|y^T)$.

\begin{example}[(Heavy-tailed nonlinear state space model)]\label{ex3}
Consider the following non-Gaussian and nonlinear state space model
\begin{eqnarray*}
(y_{t+1}|x_{t+1},\lambda_{t+1},\theta) & \sim& N(x_{t+1},\lambda
_{t+1}\sigma^2),\\
(x_{t+1}|x_t,\lambda_{t+1},\theta) &\sim& N(\beta h(x_{t}), \sigma_{x}^2),
\end{eqnarray*}
where $\theta=(\beta,\sigma^2,\tau^2)$, $h(x_{t} )=x_{t}/(1+x_{t}^{2})$
and $\lambda_{t+1} \sim \operatorname{IG}(\nu/2,\nu/2)$, for known $\nu$.
Therefore, the distribution of $(y_{t+1}|x_{t+1},\theta) \sim t_{\nu
}(x_{t+1},\sigma^2),$
that is, a\break $t$-Student with $\nu$ degrees of freedom.

\begin{figure*}[b]

\includegraphics{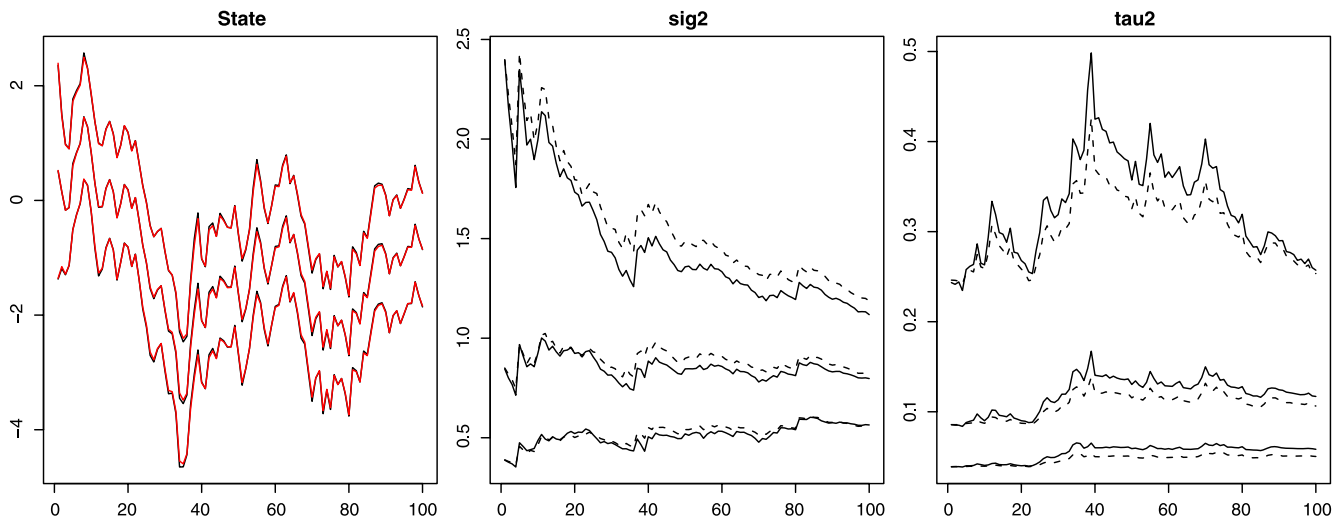}

\caption{PL and PL with state sufficient statistics (long runs).
Left panel---$p(x_t|y^t)$---PL (black), PLsuff (red);
Middle panel---$p(\sigma^2|y^t)$---PL (solid line), PLsuff (dotted line);
Right panel---$p(\tau^2|y^t)$---PL (solid line), PLsuff (dotted line).}\label{fig:lrmse}
\end{figure*}

The particle learning algorithm works as follows. Let the particle set
$\{(x_t,\theta,\lambda_{t+1},s_t)^{(i)}\}_{i=1}^N$
approximate $p(x_t,\theta,\lambda_{t+1},s_t|y^t)$. For
any
given time $t=0,\ldots,\break T-1$
and $i=1,\dots,N$, we first draw an index $k^i \sim\operatorname{Mult}(w_t^{(1)},\ldots,w_t^{(N)})$,
with $w_t^{(j)} \propto p(y_{t+1}|(x_{t},\lambda_{t+1},\break\theta)^{(j)})$,
$j=1,\ldots,N$, and
$p(y_{t+1}|x_{t},\lambda_{t+1},\theta)=\break f_N(y_{t+1};\beta
h(x_{t}),\lambda_{t+1}\sigma^{2} + \tau^2)$.
Then, we draw\vspace*{1pt} a new state\vspace*{1pt} $x_{t+1}^{(i)} \sim p(x_{t+1}| ( \lambda
_{t+1} ,
x_{t} , \theta)^{(k^i)}, y_{t+1}) \equiv f_N(x_{t+1};\break\mu
_{t+1}^{(i)},V_{t+1}^{(i)})$, where
$\mu_{t+1}=V_{t+1}(\lambda_{t+1}^{-1} \sigma^{-2} y_{t+1} + \tau
^{-2}\cdot\beta h(x_{t}))$ and
$V_{t+1}^{-1}=\lambda_{t+1}^{-1} \sigma^{-2} + \tau^{-2}$.
Finally,\vspace*{1pt} similar to Example \ref{ex1}, posterior parameter learning for $\theta
= (\beta,\sigma^2,\tau^2)$ follows directly from a conditionally
normal-inverse gamma update. Figure \ref{fig:fig3} illustrates the
above PL algorithm in a simulated example where $\beta=0.9$, $\sigma
^2=0.04$ and $\sigma^2_x = 0.01$. The algorithm uncovers the true
parameters very efficiently in a sequential fashion.
In Section \ref{sec:MCMC} we revisit this example to compare the
performances of PL,
MCMC (Carlin, Polson and Stoffer, \citeyear{CPS1992}) and the benchmark particle
filter with parameter
learning (Liu and West, \citeyear{LW2001}).\vspace*{-3pt}
\end{example}

\section{Comparing Particle Learning to Existing Methods}\label{sec6}
We now present a series of examples that illustrate the performance of
PL benchmarked by commonly used alternatives.

\begin{figure*}[t]

\includegraphics{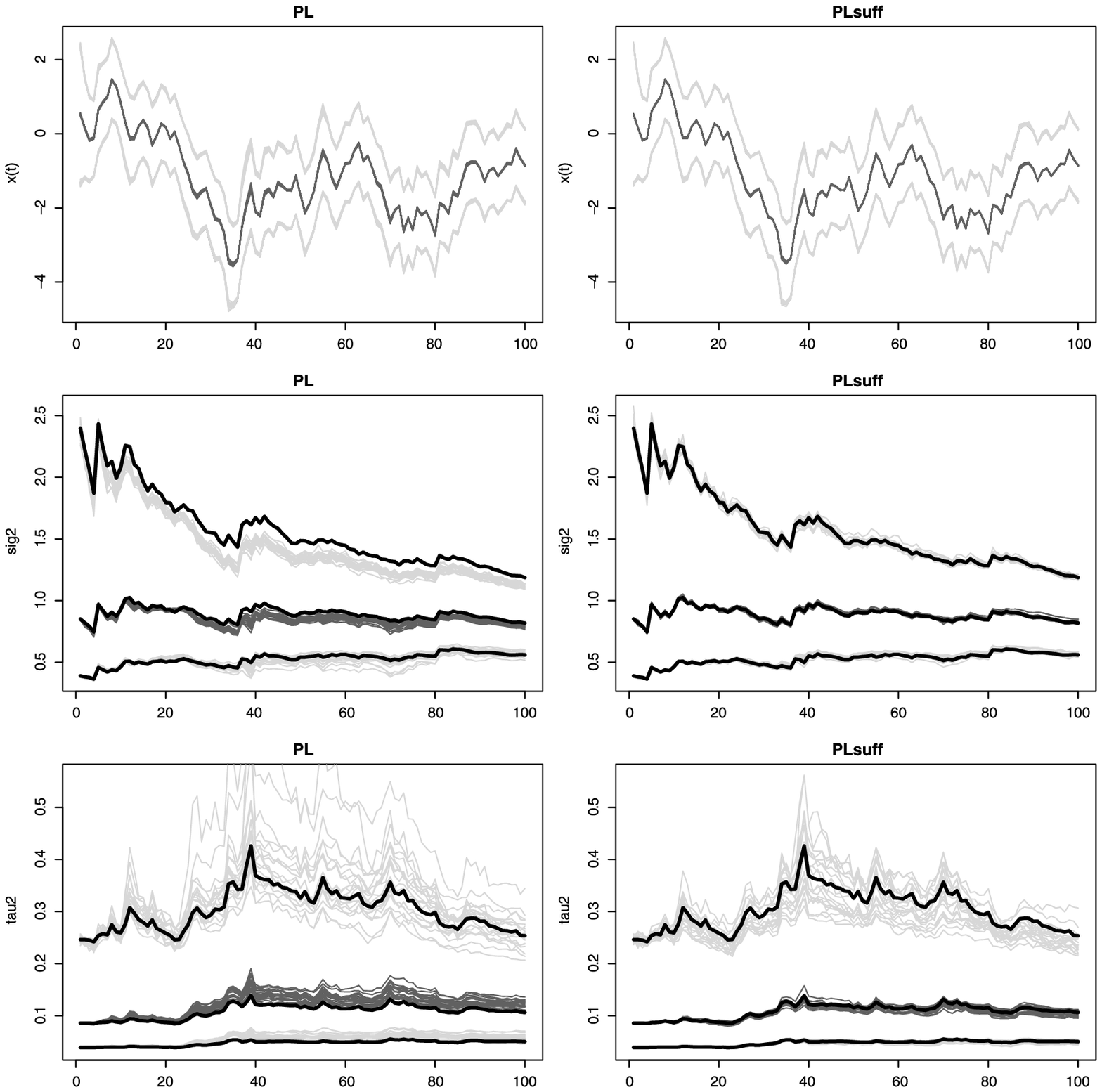}

\caption{PL and PL with state sufficient statistics (20 short runs).
PL runs (left columns) and PLsuff runs (right columns).
One long run (black) and 20 short runs (gray);
$p(x_t|y^t)$ (top row), $p(\sigma^2|y^t)$ (middle row) and $p(\tau^2|y^t)$ (bottom row).}\label{fig:lrmse1}
\end{figure*}

\begin{figure*}

\includegraphics{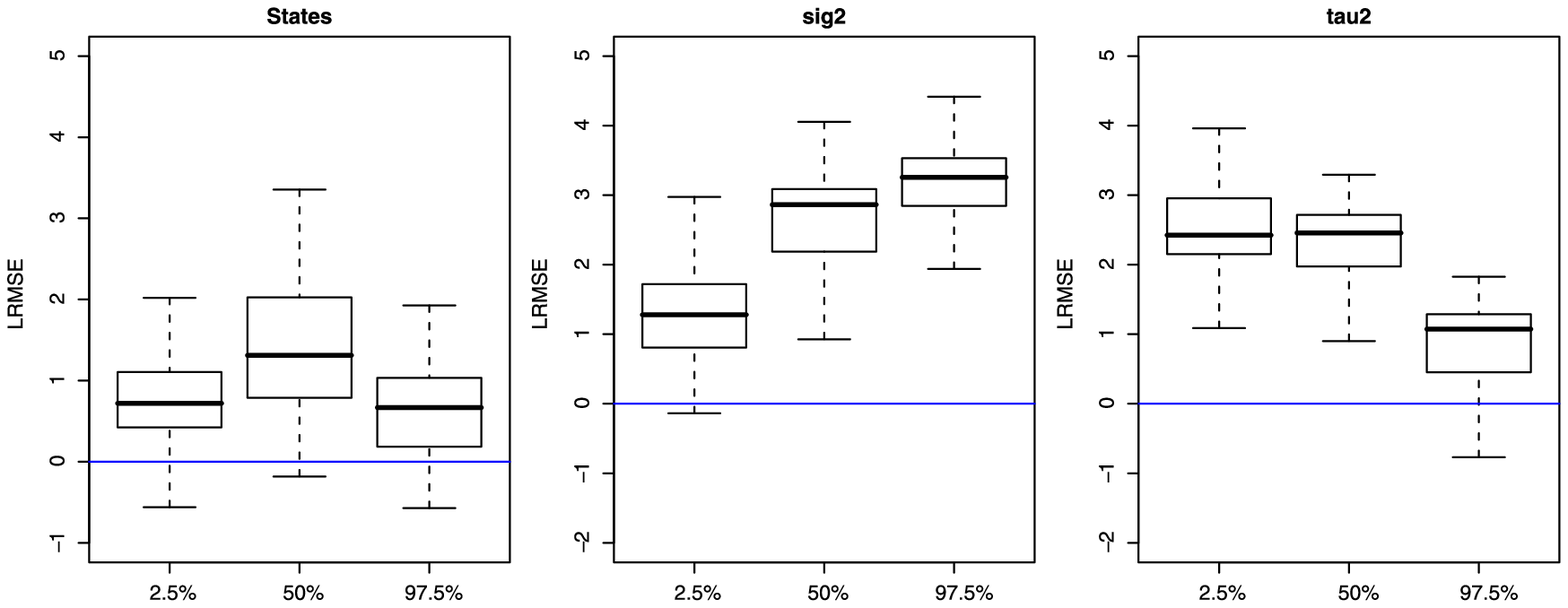}

\caption{PL and PL with state sufficient statistics (mean square errors).
Logarithm of the relative\vspace*{1pt}
mean square error for three quantiles of $p^N(x_t|y^t)$, $p^N(\sigma^2|y^t)$
and $p^N(\tau^2|y^t)$, averaged
across the 20 $N=5000$ runs.  PL relative to PLsuff.}\label{fig:lrmse2}
\end{figure*}

\begin{example}[(State sufficient statistics)]\label{ex4}
In this first simulation exercise we revisit the local level model of
Example \ref{ex1} in order to compare PL to its version that takes advantage of
state sufficient statistics, that is, by marginalizing the latent
states. The main goal is to study the Monte Carlo error of the two filters.
We simulated a time series of length $T=100$ with $\sigma^2=1$, $\tau
^2=0.1$ and $x_0=0_p$. The prior distributions are
$\sigma^2 \sim \operatorname{IG}(5,4)$, $\tau^2 \sim \operatorname{IG}(5,0.4)$ and $x_0 \sim N(0,10)$.
We run two filters: one with sequential learning for $x_t$, $\sigma^2$
and $\tau^2$
(we call it simply \textit{PL}), and the other with sequential learning
for state sufficient statistics, $\sigma^2$ and $\tau^2$ (we call it
\textit{PLsuff}).
In both cases, the particle filters are based on either one long
particle set of size $N=100\mbox{,}000$ (we call it \textit{Long})
or 20 short particle sets of size $N=5000$ (we call it \textit{Short}).
The results are in Figures \ref{fig:lrmse}
to \ref{fig:lrmse2}. Figure \ref{fig:lrmse} shows that the differences
between \textit{PL} and \textit{PLsuff} dissipate
for fairly large $N$. However, when $N$ is small \textit{PLsuff} has
smaller Monte Carlo error and is
less biased than \textit{PL}, particularly when estimating $\sigma^2$ and
$\tau^2$ (see Figure \ref{fig:lrmse1}).
Similar findings appear in Figure \ref{fig:lrmse2} where the mean
square errors of the quantiles from the 20 \textit{Short} runs are
compared to those from the \textit{Long} \textit{PLsuff} run.
\end{example}

\begin{example}[(Resample--propagate or propagate--resample?)]\label{ex5}
In this second simulation exercise we continue focusing in the local
level model of Example~\ref{ex1} to
compare PL to three other particle filters:
the bootstrap filter (BF), its fully adapted version (FABF), and
the auxiliary particle filter (APF) (no fully adapted).
BF and FABF are propagate--resample filters, while PL and APF are
resample--propagate filters.
The main goal is to study the Monte Carlo error of the four filters.
We start with the pure case scenario, that is, with fixed parameters.
We simulated 20 time series
of length $T=100$ from the local level model with parameters $\tau
^2=0.013$, $\sigma^2=0.13$ and $x_0=0$.
Therefore, the signal to noise ratio $\sigma_x/\sigma$ equals $0.32$.
Other combinations were also tried and similar results were found.
The prior distribution of the initial state $x_0$ was set at $N(0,10)$.
For each time series, we run 20 times
on each of the four filters, all based on $N=1000$ particles.
We use five quantiles to compare the various filters. Let $q_\alpha^t$
be such that
$\operatorname{Pr}(x_t < q^\alpha_t|y^t) = \alpha$, for $\alpha=(0.05,0.25,0.5,0.75,0.95)$.
Then, the mean square error (MSE) for filter $f$, at time $t$ and
quantile $\alpha$ is
\[
\mathit{MSE}^{\alpha}_{t,f} = \frac{1}{400} \sum_{d=1}^{20}\sum_{r=1}^{20}
(q^\alpha_{t,d} - {\hat q}^\alpha_{t,d,f,r})^2,
\]
where $d$ and $r$ index the data set and the particle filter run, respectively.
We compare PL, APF and FABF via logarithm relative MSE (LRMSE),
relative to the benchmark BF.
Results are summarized in Figure \ref{fig:dlm-simg2}. PL is uniformly
better than all three alternatives.
Notice that the only algorithmic difference between PL and FABF is that
PL reverses the propagate--resample steps.

\begin{figure*}[b]

\includegraphics{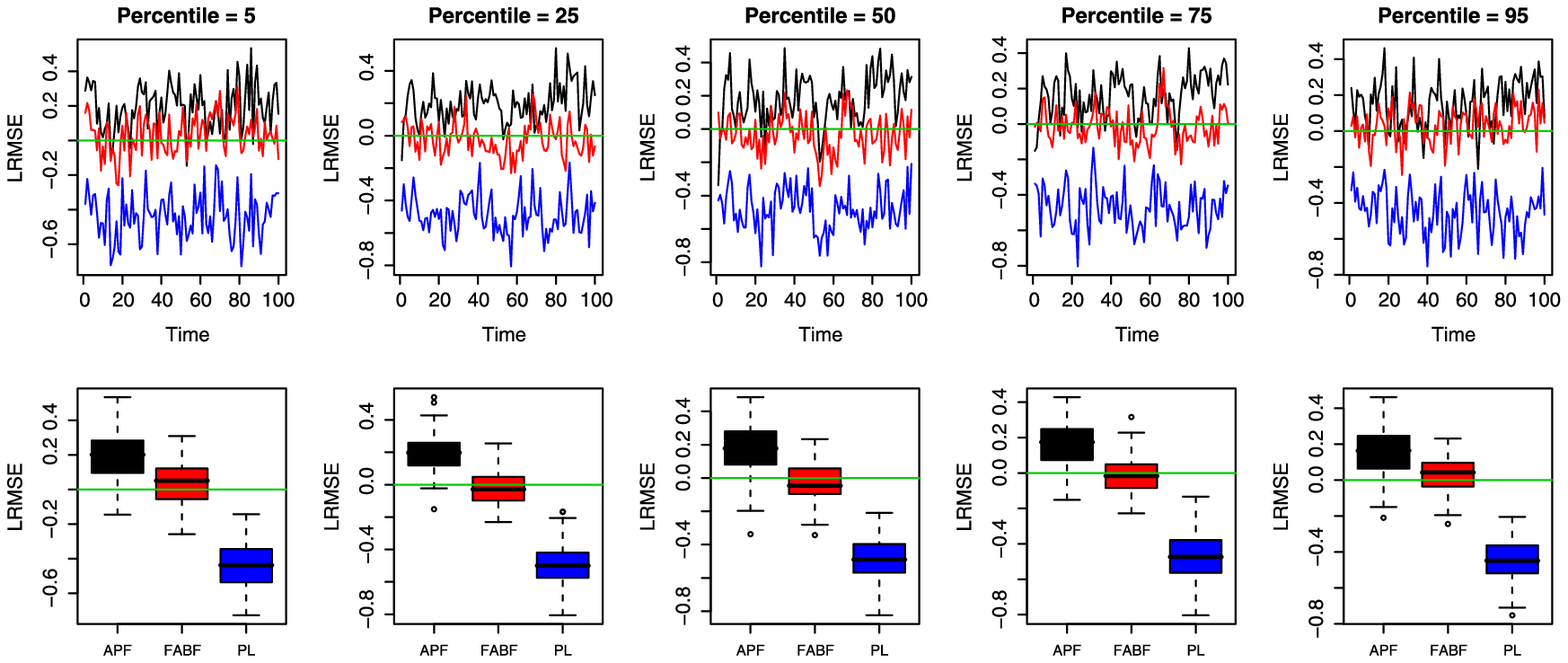}

\caption{APF, FABF and PL pure filter.  Logarithm of the relative
mean square error for five quantiles of $p^N(x_t|y^t)$.  MSE relative to BF.
Boxplots on the second row are based on the
time series plots on the first row.}\label{fig:dlm-simg2}
\end{figure*}

We now move to the parameter learning scenario, where $\sigma^2$ is
still kept fixed but learning
of $\tau^2$ is performed. Three time series of length $T=1000$ were
simulated from the local level model
with $x_0=0$ and $(\sigma^2,\tau^2)$ in $\{
(0.1,0.01),(0.01,0.01),(0.01,0.1)\}$. The independent prior
distributions for $x_0$ and $\tau^2$ are $x_0 \sim N(0,1)$ and $\tau^2
\sim \operatorname{IG}(10,9\tau_0^2)$,
where $\tau_0^2$ is the true value of $\tau^2$ for a given time series.
In all filters $\tau^2$ is sampled offline from $p(\tau^2|s_t)$ where
$s_t$ is the vector of
conditional sufficient statistics. We run the filters 100 times, all
with the same seed within run, for each
one of the three simulated data sets. Finally, the number of particles
was set at $N=5000$, with similar results
found for smaller $N$, ranging from 250 to 2000 particles. Mean
absolute errors (MAE) over the 100 replications
are constructed by comparing quantiles of the true sequential
distributions $p(x_t|y^t)$ and $p(\tau^2|y^t)$ to quantiles of the
estimated sequential distributions $p^N(x_t|y^t)$ and $p^N(\tau^2|y^t)$.
More specifically, for time $t$, $a$ in $\{x,\tau^2\}$, $\alpha$ in $\{
0.01,0.50,0.99\}$, true
quantiles $q_{t,a}^\alpha$ and PL quantiles ${\hat q}_{t,a,r}^\alpha$,
\[
\mathit{MAE}_{t,a}^\alpha= \frac{1}{100}\sum_{r=1}^{100} |q_{t,a}^\alpha-{\hat
q}_{t,a,r}^\alpha|.
\]
Across different quantiles and combinations of error variances, PL is
at least as good as FABF and in many cases significantly better than
BF. Results appear in Figure \ref{fig:5000tau2}.
\end{example}

\begin{figure*}

\includegraphics{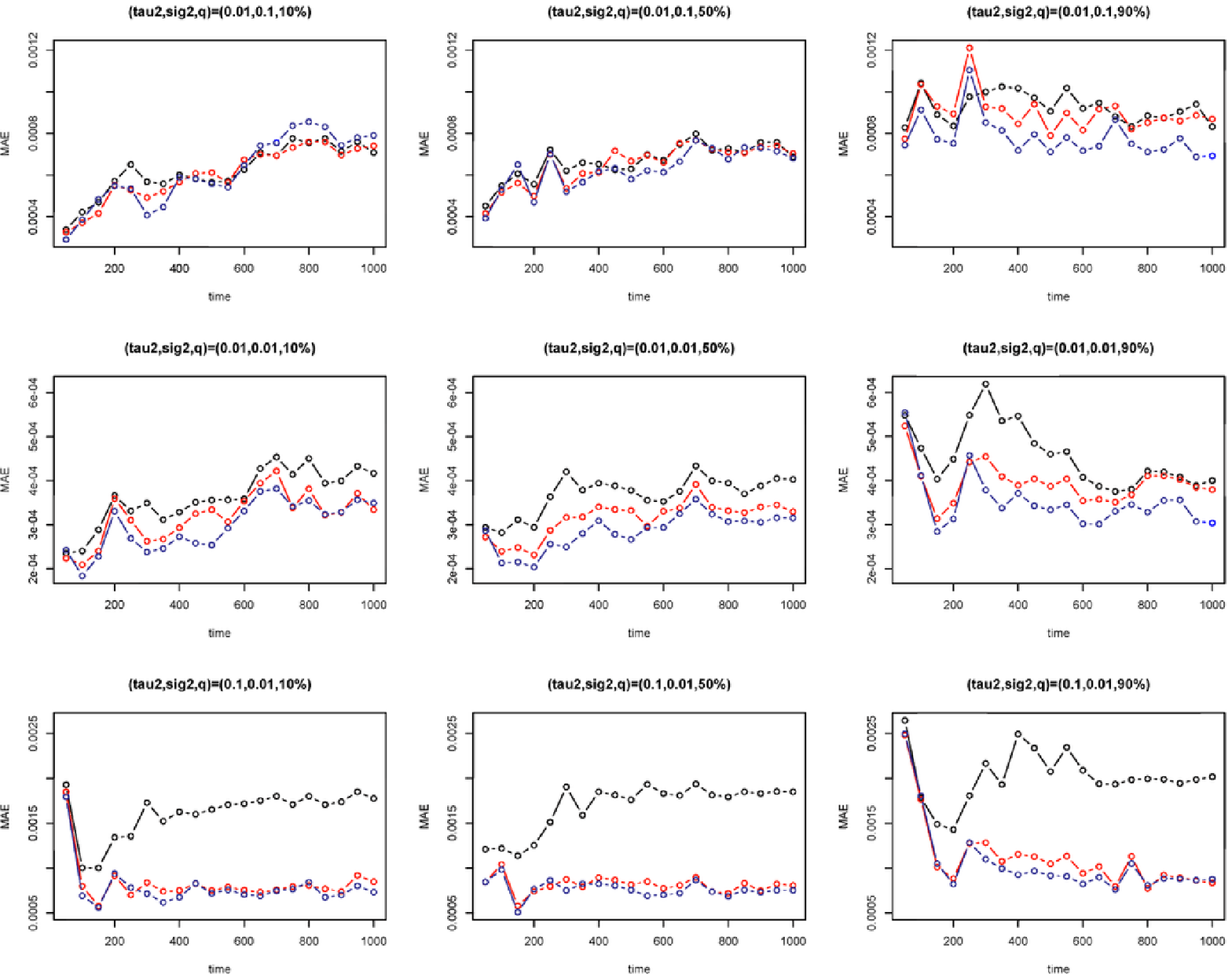}

\caption{BF, FABF and PL with learning of $\tau^2$. Mean absolute errors.
BF (black), FABF (red) and PL (blue).}\label{fig:5000tau2}
\end{figure*}

\begin{figure*}[t]

\includegraphics{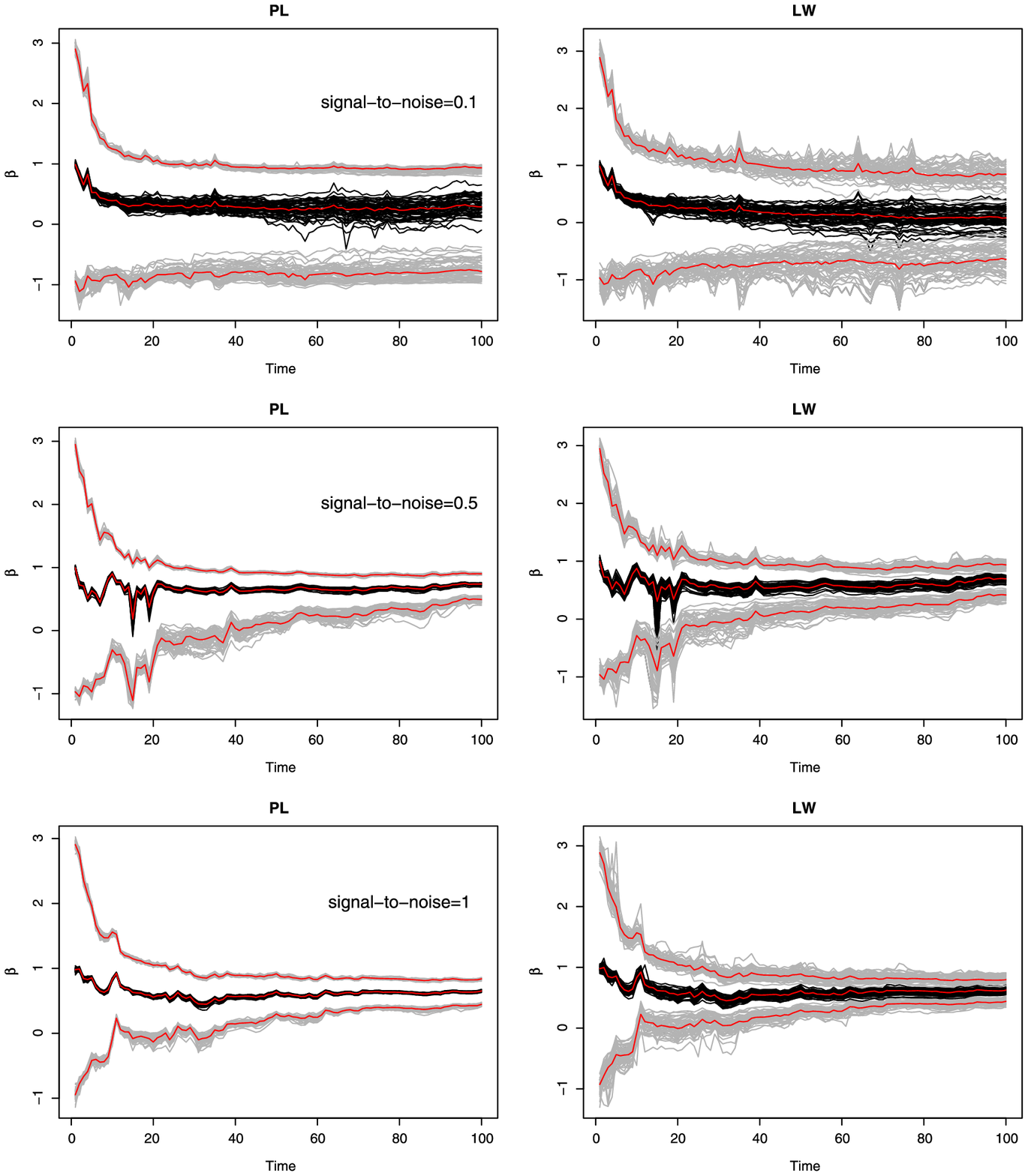}

\caption{PL and LW (parameter learning).
Posterior mean and 95\% credibility interval from $p(\beta|y^t)$.
Medians across the 50 runs appear in red.
$N=2000$ particles.  \texttt{signal-to-noise} stands for $\sigma_x/\sigma$.
In all cases, $\sigma=1$.}\vspace*{-2pt}
\label{fig:lw}
\end{figure*}

\begin{example}[(PL versus LW)]\label{ex6}
Consider once again a variation of the dynamic linear model introduced
in Example \ref{ex1}, but now we assume complete knowledge
about $(\sigma^2,\tau^2)$ in
\begin{eqnarray*}
(y_{t+1}|x_{t+1},\beta) &\sim& N(x_t,\sigma^2),\\
(x_{t+1}|x_t,\beta) &\sim& N(\beta x_t,\tau^2)
\end{eqnarray*}
for $t=1,\ldots,T=100$, $\sigma^2=1$, $x_1=0.0$ and three possible
values for $\tau^2=(0.01,0.25,1.00)$.
So, the signal to noise ratio $\tau/\sigma=0.1,0.5,1.0$. Only $\beta$
and $x_t$ are sequentially estimated and their independent prior
distributions are $N(1.0,1.0)$ and $N(0.0,1.0)$, respectively.
The particle set has length $N=2000$ and both filters were run $50$
times to study the size of the Monte Carlo error. The smoothing
parameter $\delta$ of Liu and West's filter was set at $\delta=0.95$,
but fairly similar results were found for $\delta$ ranging from $0.8$
to $0.99$.
Our findings, summarized in Figure \ref{fig:lw}, favor PL over LW
uniformly across all scenarios. The discrepancy is higher when
$\tau/\sigma$ is small, which is usually the case in state space applications.
\end{example}

\begin{figure*}

\includegraphics{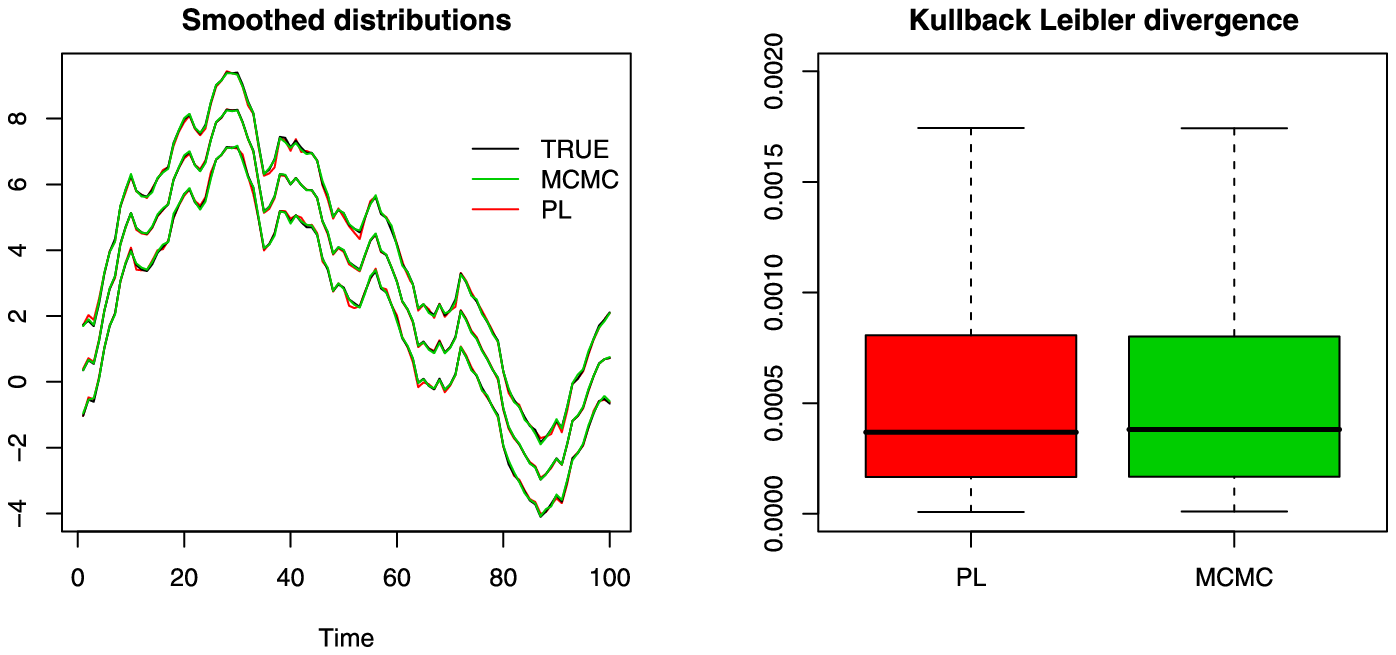}

\caption{PL and FFBS (smoothed distributions).
$T=100$ simulated from a local level model with $\sigma^2=1$, $\tau^2=0.5$, $x_0=0$ and
$x_0 \sim N(0,100)$.  PL is based on $N=1000$ particles, while FFBS is based on $2N$ draws with
the first $N$ discarded.}\label{fig:smoothing}\vspace*{-8pt}
\end{figure*}

\subsection{PL vs MCMC}
\label{sec:MCMC}
PL combined with the backward smoothing algorithm (as in Section \ref{sec:smooth})
is an alternative to MCMC methods for state space models.
In general, MCMC methods (see Gamerman and Lopes, \citeyear{GL2006}) use\break Markov
chains designed to explore the posterior distribution
$p(x^T,\theta|y^T) $ of states and parameters
conditional on all the information available, $y^T = ( y_1 ,
\ldots,\break
y_T )$. For example, an MCMC strategy would have to iterate through
\[
p(\theta|x^T,y^T) \quad \mbox{and}\quad  p(x^T|\theta, y^T).
\]
However, MCMC relies on the convergence of very
high-dimensional Markov chains. In the purely conditional Gaussian
linear models or when states are dicrete,
$p(x^T|\theta,y^T)$ can be sampled in block using FFBS. Even in these ideal
cases, achieving convergency is far from an easy task and the computational
complexity is enormous, as at each iteration one would have to filter forward
and backward sample for the full state vector $x^T$. The particle
learning algorithm
presented here has two advantages: (i) it requires only one
forward/backward pass through the data
for all $N$ particles and (ii) the approximation accuracy does not
rely on
convergence results that are virtually impossible to assess in practice
(see Papaspiliopoulos and Roberts, \citeyear{PR2008}).

In the presence of nonlinearities, MCMC methods will suffer even
further, as no FFBS
scheme is available for the full state vector $x^T$. One would have to
resort to univariate
updates of $p(x_{t}|x_{(-t)},\theta,y^T)$ as in Carlin, Polson and
Stoffer (\citeyear{CPS1992}), where
$x_{(-t)}$ is $x^T$ without $x_t$. It is well
known that these methods generate very ``sticky'' Markov chains, increasing
computational complexity and slowing down convergence.
PL is also attractive given the simple nature of its implementation
(especially if compared to more novel hybrid methods).

\begin{figure*}

\includegraphics{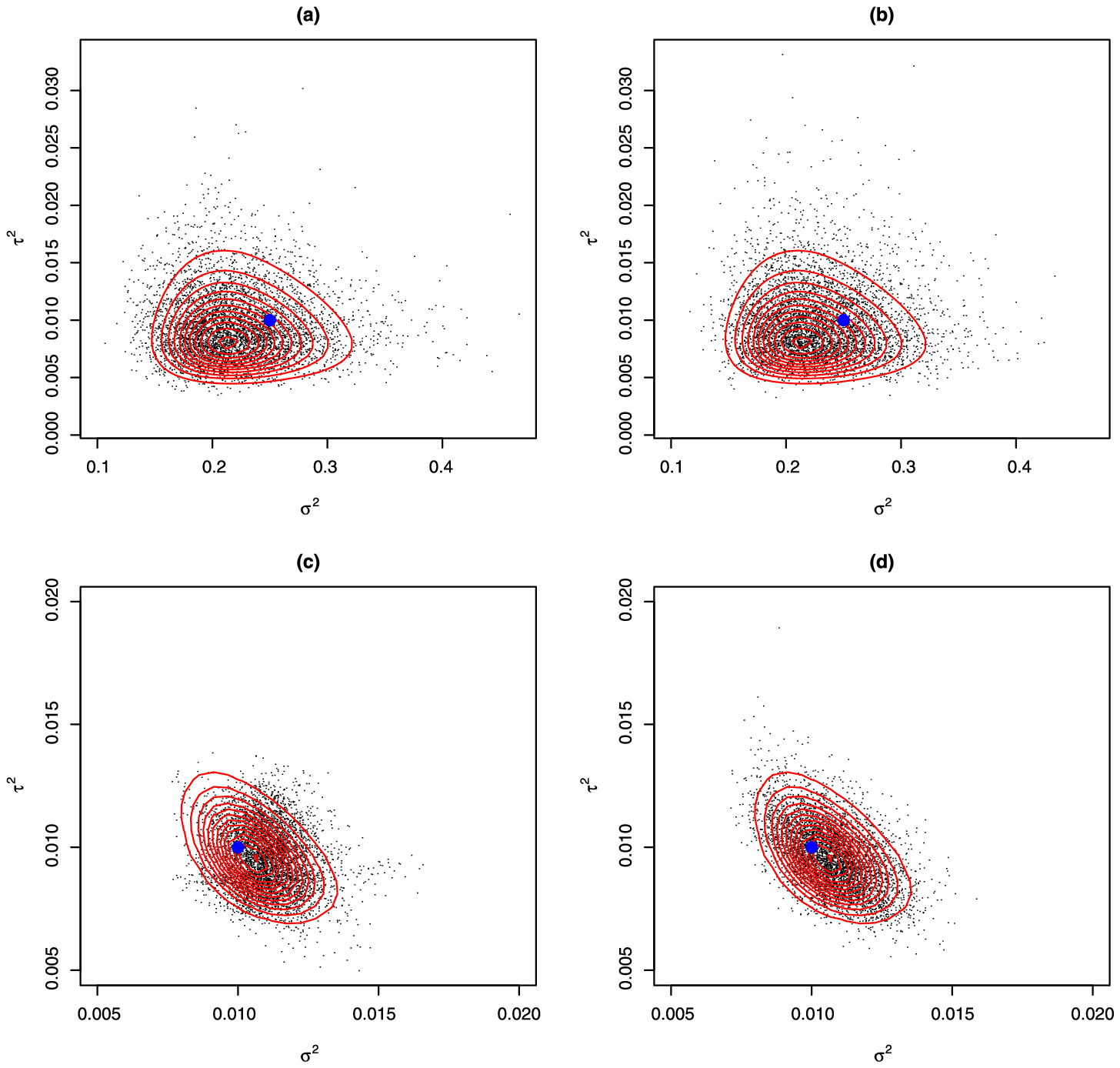}

\caption{PL and FFBS (parameter learning).
Contour plots for the true posterior $p(\sigma^2,\tau^2|y^T)$ (red contours)
and posterior draws from PL,
panels \textup{(a)} and \textup{(c)}, and FFBS, panels \textup{(b)} and \textup{(d)}.
The blue dots represent the true value of the
pair $(\sigma^2,\tau^2)$.  The sample size is $T=50$ (top row) and
$T=500$ (bottom row).}\vspace*{-8pt}
\label{fig:fig4}
\end{figure*}

\begin{example}[(PL versus FFBS)]\label{ex7}
We revisit the first order dynamic linear model introduced in Example \ref{ex1}
to compare our PL smoother and
the forward-filtering, backward-sampling (FFBS) smoother.
Assuming knowledge about $\theta$, Figure \ref{fig:smoothing} compares
the true
smoothed distributions $p(x_t|y^T)$ to approximations based on PL and
on FFBS. Now, when parameter learning is introduced, PL performance is
comparable to that of the FFBS when approximating $p(\sigma^2,\tau^2|y^T)$,
as shown in Figure \ref{fig:fig4}. We argue that, based on these
empirical findings, PL and FFBS are equivalent alternatives for
posterior computation.
We now turn to the issue of computational cost, measured here by the
running time in seconds of both schemes.
Data was simulated based on $(\sigma^2,\tau^2,x_0)=(1.0,0.5,0.0)$.
The prior distribution of $x_0$ is $N(0,100)$, while $\sigma^2$ and
$\tau^2$ are kept fixed throughout this exercise.
PL was based on $N$ particles and FFBS based on $2N$ iterations, with
the first $M$ discarded.
Table \ref{tab:time} summarizes the results.
For fixed $N$, the (computational) costs of both PL and FFBS increase
linearly with $T$, with FFBS twice as fast as PL.
For fixed $T$, the cost of FFBS increases linearly with $N$, while the
cost of PL increases exponentially with $N$.
These findings were anticipated in Section \ref{sec:smooth}. As
expected, PL outperforms FFBS when comparing filtering times.
\end{example}

\begin{example}[(PL versus single-move MCMC)]
Our final example compares PL to a single-move MCMC as in Carlin,
Polson and Stoffer (\citeyear{CPS1992}).
We consider the first order conditional Gaussian dynamic model with
nonlinear state equation as defined in Example \ref{ex3}. The example focuses
on the estimation of $\sigma^2$. We generate data with different levels
of signal to noise ratio and compare the performance of PL versus MCMC.
Table \ref{tab:tab2} presents the results for the comparisons. Once
again, PL provides significant improvements in computational
time and MC variability for parameter estimation over MCMC.
\end{example}

%
\begin{table}[t]\vspace*{6pt}
\caption{Computing time (in seconds) of PL and FFBS for smoothing. In
parenthesis are PL times for filtering}\label{tab:time}
\begin{tabular*}{\columnwidth}{@{\extracolsep{4in minus 4in}}lccccc@{}}
\hline
\multicolumn{3}{@{}c}{$\bolds{N=500}$} & \multicolumn{3}{c@{}}{$\bolds{T=100}$} \\
\ccline{1-3,4-6}
$\bolds{T}$ & \textbf{PL} & \textbf{FFBS} & $\bolds{N}$ & \textbf{PL} & \textbf{FFBS} \\
\hline
\hphantom{0}200 & 18.8 (0.25) & \hphantom{0}9.1 & \hphantom{0}500 & \hphantom{00}9.3 (0.09)& \hphantom{0}4.7 \\
\hphantom{0}500 & 47.7 (1.81) & 23.4 &1000 & \hphantom{0}32.8 (0.15)& \hphantom{0}9.6\\
1000 & 93.9 (8.29) & 46.1& 2000 & 127.7 (0.34) & 21.7\\
\hline
\end{tabular*}
\end{table}
%
\begin{table}
\caption{Single-move MCMC based on 2000 draws, after 2000 burn-in. PL
based on 2000 particles.
Expectations are with respect to the whole data set at time $T$, while
the true value of $\tau^2$ is $0.01$.
Numbers in parenthesis are 1000 times the standard deviation based on
20 replications of the algorithms.
Time is in seconds when running our code in R version 2.8.1 on a
MacBook with a 2.4 GHz processor and 4 GB MHz of memory}
\label{tab:tab2}
\begin{tabular*}{\columnwidth}{@{\extracolsep{\fill}}lcccc@{}}
\hline
$\bolds{T}$ & $\bolds{\sigma^2}$ & \textbf{Time} & $\bolds{E(\sigma^2)}$ & $\bolds{E(\tau^2)}$ \\
\hline
\multicolumn{5}{@{}c@{}}{Single-move MCMC}\\
\hphantom{0}50&0.2500&19.7&0.209934 (3.901)&0.011 (1.532)\\
&0.0100&19.3&0.009151 (0.253)&0.008 (0.545)\\
&0.0001&19.3&0.000097 (0.003)&0.010 (0.049)\\[3pt]
200&0.2500&79.3&0.249059 (6.981)&0.027 (12.76)\\
&0.0100&79.1&0.009740 (0.305)&0.013 (1.375)\\
&0.0001&79.8&0.000099 (0.004)&0.011 (0.032)\\[6pt]
\multicolumn{5}{@{}c@{}}{PL}\\
\hphantom{0}50&0.2500&\hphantom{0}0.8&0.170576 (1.633)&0.010 (0.419)\\
&0.0100&\hphantom{0}0.7&0.007204 (0.151)&0.008 (0.165)\\
&0.0001&\hphantom{0}0.6&0.000092 (0.004)&0.010 (0.058)\\[3pt]
200&0.2500&\hphantom{0}6.5&0.262396 (6.392)&0.009 (1.332)\\
&0.0100&\hphantom{0}6.4&0.010615 (0.570)&0.011 (0.935)\\
&0.0001&\hphantom{0}6.4&0.000098 (0.010)&0.011 (0.057)\\
\hline
\end{tabular*}
\end{table}

\section{Final Remarks}
\label{sec:conclusion}
In this paper we provide particle learning tools (PL) for a large class
of state space models.
Our methodology incorporates sequential parameter learning, state
filtering and smoothing.
This provides an alternative to the popular FFBS/MCMC (Carter and Kohn, \citeyear{CK1994})
approach for conditional dynamic linear models (DLMs) and also to MCMC
approaches
to nonlinear non-Gaussian models. It is also a generalization of the
mixture Kalman filter (MKF) approach of Chen and Liu (\citeyear{CL2000})
that includes parameter learning and smoothing.
The key assumption is the existence of a conditional sufficient
statistic structure for the
parameters which is commonly available in many commonly used
models.

We provide extensive simulation evidence to address the efficiency of PL
versus standard methods. Computational time and accuracy are used to
assess the
performance. Our approach compares very favorably with these existing
strategies and is robust to particle degeneracies as the sample size
grows. Finally, PL has the additional advantage of being an intuitive
and easy-to-implement computational scheme and should, therefore,
become a default choice for posterior inference in a variety of models,
with examples already appearing in Lopes et al. (\citeyear{Letal2010}), Carvalho et al. (\citeyear{Cetal2009}), Prado and Lopes
(\citeyear{PL2010}), Lopes and Tsay (\citeyear{LT2010}) and Lopes and Polson (\citeyear{LP2010}).

\section*{Acknowledgments}
We thank the Editor, Raquel Prado, Peter M\"uller and Mike West for
their invaluable comments that
greatly improved the presentation of the ideas of the paper. R~code for
all examples are freely
available upon request.


\begin{thebibliography}{99}

\bibitem[\protect\citeauthoryear{}{2010}]{BDM2010}
\textsc{Briers, M., Doucet, A.} and \textsc{Maskell, S.} (2010).
Smoothing algorithms for state-space models.
\textit{Ann. Inst. Statist. Math.} \textbf{62} 61--89.
\MR{2577439}

\bibitem[\protect\citeauthoryear{}{2007}]{CGM2007}
\textsc{Capp\'e, O., Godsill, S.} and \textsc{Moulines, E.} (2007).
An overview of existing methods and recent advances in sequential Monte Carlo.
\textit{IEEE Proceedings} \textbf{95} 899--924.

\bibitem[\protect\citeauthoryear{}{1992}]{CPS1992}
\textsc{Carlin, B., Polson, N. G.} and \textsc{Stoffer, D.} (1992).
A Monte Carlo approach to nonnormal and nonlinear state-space modeling.
\textit{J. Amer. Statist. Assoc.} \textbf{87} 493--500.

\bibitem[\protect\citeauthoryear{}{1994}]{CK1994}
\textsc{Carter, C.} and \textsc{Kohn, R.} (1994).
On Gibbs sampling for state space models.
\textit{Biometrika} \textbf{82} 339--350.
\MR{1311096}

\bibitem[\protect\citeauthoryear{}{2007}]{CL2007}
\textsc{Carvalho, C. M.} and \textsc{Lopes, H. F.} (2007).
Simulation-based sequential analysis of Markov switching stochastic
volatility models.
\textit{Comput. Statist. Data Anal.} \textbf{51} 4526--4542.
\MR{2364463}

\bibitem[\protect\citeauthoryear{}{2009}]{Cetal2009}
\textsc{Carvalho, C. M., Lopes, H. F., Polson, N. G.} and \textsc
{Taddy, M.} (2009).
Particle learning for general mixtures.
Working paper, Univ. Chicago Booth School of Business.

\bibitem[\protect\citeauthoryear{}{2000}]{CL2000}
\textsc{Chen, R.} and \textsc{Liu, J.} (2000).
Mixture Kalman filters.
\textit{J. Roy. Statist. Soc. Ser. B} \textbf{62} 493--508.
\MR{1772411}

\bibitem[\protect\citeauthoryear{}{2001}]{DFG2001}
\textsc{Doucet, A., de Freitas, J.} and \textsc{Gordon, N.} (2001).
\textit{Sequential Monte Carlo Methods in Practice}.
Springer, New York.
\MR{1847783}

\bibitem[\protect\citeauthoryear{}{2002}]{F2002}
\textsc{Fearnhead, P.} (2002).
Markov chain Monte Carlo, sufficient statistics, and particle filters.
\textit{J. Comput. Graph. Statist.} \textbf{11} 848--862.
\MR{1951601}

\bibitem[\protect\citeauthoryear{}{2008}]{FWT2008}
\textsc{Fearnhead, P., Wyncoll, D.} and \textsc{Tawn, J.} (2008).
A sequential smoothing algorithm with linear computational cost.
Working paper, Dept. Mathematics and Statistics, Lancaster Univ.

\bibitem[\protect\citeauthoryear{}{1994}]{FS1994}
\textsc{Fr\"uhwirth-Schnatter, S.} (1994).
Applied state space modelling of non-Gaussian time series using
integration-based Kalman filtering.
\textit{Statist. Comput.} \textbf{4} 259--269.

\bibitem[\protect\citeauthoryear{}{2006}]{GL2006}
\textsc{Gamerman, D.} and \textsc{Lopes, H. F.} (2006).
\textit{Markov Chain Monte Carlo: Stochastic Simulation for Bayesian Inference}.
Chapman \& Hall/CRC Press, Boca Raton, FL.
\MR{2260716}

\bibitem[\protect\citeauthoryear{}{2004}]{GDW2004}
\textsc{Godsill, S. J., Doucet, A.} and \textsc{West, M.} (2004).
Monte Carlo smoothing for nonlinear time series.
\textit{J. Amer. Statist. Assoc.} \textbf{99} 156--168.
\MR{2054295}

\bibitem[\protect\citeauthoryear{}{1993}]{GSS1993}
\textsc{Gordon, N., Salmond, D.} and \textsc{Smith, A. F. M.} (1993).
Novel approach to nonlinear/non-Gaussian Bayesian state estimation.
\textit{IEE Proceedings-F} \textbf{140} 107--113.


\bibitem[\protect\citeauthoryear{}{2008}]{JP2008}
\textsc{Johannes, M.} and \textsc{Polson, N. G.} (2008).
Exact particle filtering and learning.
Working paper, Univ. Chicago Booth School of Business.

\bibitem[\protect\citeauthoryear{}{2008}]{JPY2008}
\textsc{Johannes, M., Polson, N. G.} and \textsc{Yae, S. M.} (2008).
Nonlinear filtering and learning.
Working paper, Univ. Chicago Booth School of Business.

\bibitem[\protect\citeauthoryear{}{1960}]{K1960}
\textsc{Kalman, R. E.} (1960).
A new approach to linear filtering and prediction problems.
\textit{Transactions of the ASME---Journal of Basic Engineering} \textbf
{82} 35--45.

\bibitem[\protect\citeauthoryear{}{1998}]{LC1998}
\textsc{Liu, J.} and \textsc{Chen, R.} (1998). Sequential Monte Carlo
methods for dynamic systems.
\textit{J. Amer. Statist. Assoc.} \textbf{93} 1032--1044.
\MR{1649198}

\bibitem[\protect\citeauthoryear{}{2001}]{LW2001}
\textsc{Liu, J.} and \textsc{West, M.} (2001).
Combined parameters and state estimation in simulation-based filtering.
In \textit{Sequential Monte Carlo Methods in Practice}
(A. Doucet, N. de Freitas and N. Gordon, eds.).
Springer, New York.
\MR{1847793}

\bibitem[\protect\citeauthoryear{}{2010}]{LP2010}
\textsc{Lopes, H. F.} and \textsc{Polson, N. G.} (2010).
Extracting SP500 and NASDAQ volatility: The credit crisis of 2007--2008.
In \textit{Handbook of Applied Bayesian Analysis} (A. O'Hagan and M.
West, eds.) 319--342. Oxford Univ. Press, Oxford.

\bibitem[\protect\citeauthoryear{}{2010}]{LT2010}
\textsc{Lopes, H. F.} and \textsc{Tsay, R. E.} (2010).
Bayesian analysis of financial time series via particle filters.
\textit{J. Forecast.} To appear.

\bibitem[\protect\citeauthoryear{}{2010}]{Letal2010}
\textsc{Lopes, H. F., Carvalho, C. M., Johannes, M.} and \textsc
{Polson, N. G.} (2010).
Particle learning for sequential Bayesian computation. In
\textit{Bayesian Statistics 9} (J. M. Bernardo, M. J. Bayarri, J.~O.
Berger, A. P. Dawid,
D. Heckerman, A. F. M. Smith and M. West, eds.). Oxford Univ. Press, Oxford.

\bibitem[\protect\citeauthoryear{}{2008}]{PR2008}
\textsc{Papaspiliopoulos, O.} and \textsc{Roberts, G.} (2008).
Stability of the Gibbs sampler for Bayesian hierarchical models.
\textit{Ann. Statist.} \textbf{36} 95--117.
\MR{2387965}

\bibitem[\protect\citeauthoryear{}{1999}]{PS1999}
\textsc{Pitt, M.} and \textsc{Shephard, N.} (1999).
Filtering via simulation: Auxiliary particle filters.
\textit{J. Amer. Statist. Assoc.} \textbf{94} 590--599.
\MR{1702328}

\bibitem[\protect\citeauthoryear{}{2008}]{PSM2008}
\textsc{Polson, N. G., Stroud, J.} and \textsc{M\"uller, P.} (2008).
Practical filtering with sequential parameter learning.
\textit{J. Roy. Statist. Soc. Ser. B} \textbf{70} 413--428.
\MR{2424760}

\bibitem[\protect\citeauthoryear{}{2010}]{PL2010}
\textsc{Prado, R.} and \textsc{Lopes, H. F.} (2010).
Sequential parameter learning and filtering in structured
autoregressive models.
Working paper, Univ. Chicago Booth School of Business.

\bibitem[\protect\citeauthoryear{}{2002}]{S2002}
\textsc{Storvik, G.} (2002).
Particle filters in state space models with the presence of unknown
static parameters.
\textit{IEEE Trans. Signal Process.} \textbf{50} 281--289.

\bibitem[\protect\citeauthoryear{}{1986}]{W1986}
\textsc{West, M.} (1986).
Bayesian model monitoring.
\textit{J. Roy. Statist. Soc. Ser. B} \textbf{48} 70--78.
\MR{0848052}

\bibitem[\protect\citeauthoryear{}{1997}]{WH1997}
\textsc{West, M.} and \textsc{Harrison, J.} (1997).
\textit{Bayesian Forecasting and Dynamic Models}, 2nd ed.
Springer, New York.
\MR{1482232}\vspace*{-2pt}

\end{thebibliography}
\end{document}